\documentclass[apj]{emulateapj}

\shorttitle{Hot halo gas in NGC~5746 and NGC~5170}
\shortauthors{Rasmussen et al.}

\begin{document}

\title{Investigating hot gas in the halos of two massive spirals: \\
  Observations and cosmological simulations}


\author{Jesper Rasmussen,\altaffilmark{1} Jesper
  Sommer-Larsen,\altaffilmark{2} Kristian Pedersen,\altaffilmark{2}
  Sune Toft,\altaffilmark{3} Andrew Benson,\altaffilmark{4} Richard
  G.~Bower,\altaffilmark{5} and Lisbeth F.~Olsen\altaffilmark{2}}


\altaffiltext{1}{School of Physics and Astronomy, University of
  Birmingham, Edgbaston, Birmingham B15 2TT, UK;
  jesper@star.sr.bham.ac.uk}

\altaffiltext{2}{Dark Cosmology Centre, Niels Bohr Institute,
  University of Copenhagen, Juliane Maries Vej 30, DK-2100 Copenhagen,
  Denmark}

\altaffiltext{3}{European Southern Observatory, Karl-Schwarzschild-Str.\
2, 85748, Garching bei M\"{u}nchen, Germany}

\altaffiltext{4}{Division of Physics, Mathematics, \& Astronomy, California 
Institute of Technology, Mail Code 130-33, Pasadena, CA 91125, USA}

\altaffiltext{5}{Institute for Computational Cosmology, University of
  Durham, South Road, Durham DH1 3LE, UK}


\begin{abstract}
  Models of disk galaxy formation commonly predict the existence of an
  extended reservoir of hot gas surrounding massive spirals at low
  redshift.  As a test of these models, we have obtained X-ray and
  optical data of the two massive edge-on spirals NGC~5746 and
  NGC~5170, in order to investigate the amount and origin of hot gas
  in their disks and halos.  {\em Chandra} observations of NGC~5746
  reveal evidence for diffuse X-ray emission with a total luminosity
  of $\sim 7\times 10^{39}$~erg~s$^{-1}$ surrounding this galaxy out
  to at least $\sim 20$~kpc from the disk, whereas an identical study
  of the less massive NGC~5170 fails to detect any extraplanar X-ray
  emission.  Unlike the case for other disk galaxies with detected
  X-ray halos, the halo emission around NGC~5746 is not accompanied by
  extraplanar H$\alpha$ or radio emission, and there is no evidence
  for significant nuclear or starburst activity in the disk.  In
  contrast to these other cases, the emission around NGC~5746
  therefore appears to arise from the cooling of externally accreted
  material rather than from disk outflows.  To verify this idea, we
  present results of cosmological simulations of galaxy formation and
  evolution, showing our observations to be in good agreement with
  expectations for cosmological accretion, while also confirming that
  the X-ray halos of other spirals do not fit well into an accretion
  scenario. We find that the estimated cooling rate of hot halo gas
  around NGC~5746 would provide sufficient material for star formation
  in the disk to proceed at its present rate.  This lends support to
  the idea that a supply of hot ambient gas is potentially available
  as fuel for star formation in massive, nearby spirals, and suggests
  that accretion of hot gas could be important for maintaining the
  stellar disks of such galaxies. Finally, our results support the
  notion that hot halo gas constitutes most of the "missing" galactic
  baryons.
\end{abstract}


\section{Introduction}\label{sec,intro}

Estimates of the cosmic baryon fraction, defined as the ratio of
baryonic to total mass in the Universe, can be combined with
constraints on the integrated mass function of galaxies to infer that
most baryons in the Universe are in a hot, diffuse form at the present
epoch \citep{balo2001}.  Cosmological simulations suggest that
30--40~per~cent of all baryons in the Universe reside in intergalactic
filaments of shock-heated gas with temperatures $10^5\lesssim T
\lesssim 10^7$~K, the so-called warm/hot intergalactic medium, WHIM
(e.g.\ \citealt{cen1999}; \citealt{dave2001}).  X-ray absorption
studies have provided observational evidence for such a component
along the line of sight towards a number of quasars (e.g.\
\citealt{trip2000}; \citealt{nica2005}; \citealt{sava2005}).

While most of the WHIM baryons are predicted to reside in structures
of low overdensity, outside the dark matter halos of individual
galaxies and groups of galaxies \citep{dave2001}, a potential
repository for some of the ``hidden'' baryons in the Universe could be
extended halos of hot ($\sim 10^6 - 10^7$ K) gas around individual
galaxies, including spirals (e.g.\ \citealt{fuku2006}).  Cosmological
simulations suggest that the total mass of these ``external'' galactic
baryons is comparable to that of stars and cold gas in the galaxies
themselves \citep{somm2006}. In a cosmological context, observational
support for such a scenario comes from the angular correlations
between the galaxy distribution and the soft X-ray background, which
indicate the presence of soft X-ray emission from WHIM surrounding
individual galaxies \citep{solt2006}.

The presence of extended hot gaseous halos around optically bright
elliptical galaxies is well established from X-ray observations (e.g.\
\citealt{osul2001}).  The idea that also disk galaxies like the Milky
Way could be embedded in such halos dates back to the pioneering work
of \citet{spit1956} and is now integral to many semi-analytical models
of disk galaxy formation
\citep{whit1978,whit1991,kauf1993,some1999,vank1999,cole2000,hatt2003}.
In these models, galaxy dark matter halos are assumed to grow as
predicted by spherical infall models, with gas accreting continuously
along with the dark matter.  During infall into the dark matter
potential, the gas is heated, potentially to the halo virial
temperature, subsequently cooling radiatively. If cooling is rapid, as
is the case for characteristic gas temperatures $T\la 10^6$~K and
hence relatively shallow potential wells, no accretion shock develops
outside the evolving galactic disk. In these models, present-day
spirals of total mass $M\la 10^{11}$~M$_{\sun}$ accrete gas which is
predominantly in a cold, non--X-ray emitting phase at temperatures
much lower than the virial temperature of their halo
\citep{binn1977,birn2003,binn2004,kere2005}.  This 'cold accretion'
mode would be particularly pronounced at high redshift, and would
imply that most of the halo radiation is emitted as Ly$\alpha$
emission close to the disk. Indeed, observational support for this
scenario is now appearing at both low and high redshift (see, e.g.,
\citealt{vand2004,frat2006,nils2006} and references therein).

At higher galaxy masses and gas temperatures, the gas would emit at
X-ray wavelengths but would cool less efficiently, and is then assumed
in the models to flow in more slowly than in the fast-cooling regime.
A hot X-ray emitting halo can develop, in which the infalling gas is
eventually deposited where stopped by angular momentum (typically at a
central distance of a few kpc).  As in the 'cold accretion' mode, the
gas may eventually condense to form stars in a rotationally supported
disk. The accreting gas is not assumed to comprise an infinite, static
reservoir, however; the quasi-static region is assumed in the models
to be terminated by the accretion shock in the slow-cooling regime or
at the galaxy itself in the fast-cooling regime.

This general picture has in recent years been backed by some
hydrodynamical simulations of galaxy formation
\citep{toft2002,kere2005}, lending support to the general validity of
the simplifying assumptions underlying semi-analytical models. For
example, the cosmological simulations presented by \citet{toft2002}
showed the X-ray luminosity $L_{\rm X}$ of hot halo gas to scale
strongly with disk circular velocity $v_c$ (defined here as the
rotation velocity at 2.2~times the disk scale-length; see
\citealt{somm2001} for details), with $L_{\rm X} \propto v_c^{5-7}$,
as also expected from semi-analytical models of galaxy formation.
This clearly suggests that X-ray observations could provide a useful
route to mapping out any material accreting onto massive spirals at
the present epoch.

Extended X-ray halos have indeed been detected around numerous
star-forming spirals (see \citealt{stri2004};
\citealt{tull2006a,tull2006b} and references therein;
\citealt{li2006}).  However, these galaxies typically have at least
moderately high star formation rates ($\ga 1$~M$_\odot$~yr$^{-1}$),
and in every single case so far, the detected halo X-ray emission
coincides with extraplanar H$\alpha$ and radio continuum emission and
can be associated with stellar feedback activity in the disk
\citep{tull2006a}. Indeed, in many of these cases there is unambiguous
kinematic evidence for outflows perpendicular to the disk
\citep{stri2004}.  Unlike the expectation for gravitationally heated
gas, the X-ray luminosity of the gas in the halos of these galaxies is
found to correlate with various measures of the disk star formation
rate and with the energy input rate by supernovae (SN), but not with
baryonic or H{\sc i} mass in the disk.  These works suggest the
existence of a limiting star formation rate of $\approx
1$~M$_{\sun}$~yr$^{-1}$, below which such multi-phase gaseous halos
are not generated \citep{tull2006b}. The halos observed so far
therefore seem to result as a consequence of stellar feedback in the
disk rather than infalling intergalactic material.

While recent results indicate that ongoing cosmological accretion of
gas is generally necessary to explain certain properties of spiral
disks such as lopsidedness and warps (e.g.\ \citealt{bour2005};
\citealt{shen2006}) and the presence of bars
\citep{bloc2002,bour2002}, the gas accretion history of spirals in
general remains a poorly understood topic from an observational
viewpoint.  There is one case, however, the Milky Way itself, for
which indirect arguments can be used to both ascertain the presence of
a very extended hot gaseous halo (e.g.\ \citealt{moor1994};
\citealt{quil2001}; \citealt{semb2003}) and constrain the infall rate
of externally accreted gas \citep{roch1996,page1997,casu2004}.
Observations of the X-ray background also supports the idea that a
reservoir of hot virial-temperature gas surrounds the Milky Way (e.g.\
\citealt{wolf1995}; \citealt{piet1998}; \citealt{kunt2000}).  Despite
this, it is neither clear whether any of the gas in this putative hot
halo is actually cooling out to accrete onto the Milky Way disk, nor
to what extent any infall of this gas balances that currently being
removed from the disk by star formation, although some of the
high-velocity H{\sc i} clouds surrounding the Milky Way may represent
condensed, accreting material.

Hence, with the possible exception of the Milky Way, extended, gaseous
halos that can be unambiguously associated with cosmological accretion
of gas onto spirals have yet to be detected (but see below).  In a
first study dedicated to directly detecting these halos through their
X-ray emission, \citet{bens2000} compared {\em ROSAT} observations of
three massive, highly inclined spirals to predictions of simple
cooling flow models.  The failure of these authors to detect any hot
gas around the galaxies translated into upper limits on the halo X-ray
luminosity which, for their best-constrained case of NGC~2841, was
found to be a factor of 30 below their model prediction. While their
models were deliberately simplified versions of general
semi-analytical ones, the result persists that both some simple
analytical models and hydrodynamical simulations overpredict actual
halo luminosities (\citealt{gove2004}; \citealt{bens2000} and
references therein), though the exact magnitude of this discrepancy
has yet to be established.

More sensitive X-ray observations of carefully selected systems are
required to settle the question of whether hot gas halos exist around
quiescent spirals at all. This would be an important consistency check
on the assumptions underlying existing disk galaxy formation models,
and has become feasible with the current generation of X-ray
telescopes, which boasts substantial improvements in sensitivity and
spatial resolution compared to {\em ROSAT}.  Motivated by this, we
initiated a {\em Chandra} search for hot X-ray gas surrounding
massive, quiescent disk galaxies. As part of this program, the
discovery of hot halo gas around NGC~5746, with a total X-ray
luminosity of $\sim 5\times 10^{39}$~erg~s$^{-1}$, was reported by
\citet[hereafter Paper~I]{pede2006}.  In the present paper, we
elaborate on the presentation in Paper~I, and describe our
corresponding {\em Chandra} results for the less massive galaxy
NGC~5170. The motivation underlying our target selection is described
in detail, as is the reduction and analysis of X-ray and H$\alpha$
data of both galaxies. The derived constraints on hot halo properties
are also discussed in comparison to the X-ray halos seen around other
galaxies, both late- and early-type. A comprehensive discussion of the
origin and dynamical state of halo gas around NGC~5746 is then given.
In order to facilitate a discussion within the framework of current
theories of disk galaxy formation, we furthermore describe and present
results from cosmological simulations of galaxy formation, comparing
their predictions to the results for NGC~5746, NGC~5170, and other
massive spirals. On this background, we discuss the implications for
the gas accretion history of massive spirals and for models of disk
galaxy formation and evolution.

This paper is structured as follows. In \S~\ref{sec,sample} we outline
our target selection, and in \S~\ref{sec,obs} we detail the reduction
and analysis of X-ray and optical data.  Results for both galaxies are
presented in \S~\ref{sec,results} and compared to those of other
galaxies in \S~\ref{sec,prop}. The origin of hot halo gas around
NGC~5746 is investigated in \S~\ref{sec,origin}, and comparisons are
made to predictions of cosmological simulations in
\S~\ref{sec,comparison}. Our findings and conclusions are summarized
in \S~\ref{sec,summary}. A Hubble constant of $H_0=75$ km s$^{-1}$
Mpc$^{-1}$ is assumed throughout. The distances to NGC~5746 and
NGC~5170 are then 29.4 and 24.0 Mpc, respectively \citep{tull1988},
with 1~arcmin corresponding to $\sim 8.6$ and $\sim 7.0$~kpc.  Unless
otherwise stated, all errors are reported at the 68~per~cent
confidence level.

\section{Target Selection}\label{sec,sample}

The primary objective of this study is to map and characterize any
diffuse X-ray gas surrounding isolated, quiescent disk galaxies. Our
starting point was therefore the simulation results of
\citet{toft2002}, in particular the result that halo X-ray luminosity
$L_{\rm X}$ should scale strongly with disk circular velocity $v_c$.
Hence, our target sample was drawn by searching the NED and Hyperleda
databases for spiral galaxies with circular velocity $v_c \ge 280$ km
s$^{-1}$. According to the results of \citet{toft2002}, this would
ensure that the target galaxies are massive enough to retain an X-ray
halo detectable at the present day.  The expected extraplanar X-ray
luminosities of such galaxies would translate into a few hundred
ACIS-I counts for galaxies within distances of a few tens of Mpc and
for typical {\em Chandra} exposures of a few tens of ks.  We consider
this a reasonable lower limit for the detection of large-scale diffuse
emission around nearby galaxies with {\em Chandra}. A further
advantage of large $v_c$ lies in the fact that the efficiency with
which SN outflows can escape the disk and complicate the
interpretation of our results is expected to decline with increasing
disk mass.

In addition, we required our targets to meet the following criteria:
i) inclination $i\ge 80\degr$, enabling a clear view of any hot gas
above the disk plane; ii) distance $D<40$~Mpc, for sufficient angular
extent on the sky to enable a discrimination between disk and halo
gas, and to obtain a sufficiently strong X-ray signal; iii) Galactic
latitude $|b|>30\degr$, to have a low column density of absorbing
foreground Galactic hydrogen, given that any halo emission is expected
to be fairly soft and thus subject to significant absorption by
intervening material (this criterion excluded a galaxy like NGC~2613,
recently studied by \citealt{li2006}); iv) the galaxies should be
isolated (i.e.\ not be a member of a pair or a group/cluster) and show
no significant signs of interaction, so as to include only undisturbed
galaxies and avoid contamination from hot gas residing in an ambient
intragroup/-cluster medium; v) in order to minimize contamination from
outflows of hot gas, the galaxies should be quiescent, i.e.\ not show
evidence for significant starburst activity or activity related to an
active galactic nucleus (AGN). This excluded galaxies such as NGC~891,
which displays extraplanar X-ray, H$\alpha$, and radio emission
associated with stellar feedback activity (see \citealt{temp2005} and
references therein).

Imposing these strict criteria, only two galaxies remained, NGC~5746
and NGC~5170. We stress that, within our search criteria, our sample
is only as complete as the galaxy catalogues included within NED and
Hyperleda.  We also note that our selection was not restricted to
particular Hubble types, but given the $v_c\geq 280$~km~s$^{-1}$
criterion, the selection method is likely to bias against late-type
spirals. Since such galaxies are likely to display higher star
formation activity at present than early-type ones, this criterion is
in turn likely to favor relatively quiescent spirals, as desired. Of
our two galaxies, NGC~5746 is the most massive one, being an SBb
spiral with a circular velocity $v_c = 318\pm 10$~km~s$^{-1}$
(HyperLeda database).  For our other target, the Sc spiral NGC~5170,
updated values of $v_c$ became available after we initiated this
study, with $v_c$ now estimated at the significantly lower value of
$v_c = 247\pm3$~km~s$^{-1}$ (e.g.\ \citealt{kreg2004}) rather than the
value of $v_c=296$ km s$^{-1}$ \citep{tull1988} assumed when selecting
our targets.  We adopt here this new value.  Assuming $L_{\rm X}
\propto v_c^{5-7}$, this would imply a decrease in its predicted hot
halo luminosity by a factor of 3--4 relative to our initial
expectations.  As we will see, this suggests that we should not expect
the halo of this galaxy to be detectable in our {\em Chandra} data, a
suggestion confirmed by our analysis.  We can therefore effectively
use the data of this galaxy as a consistency check of our analysis
procedure for NGC~5746.

In terms of their star formation rate (SFR), both galaxies are fairly
quiescent, as already evidenced by the fact that none of them is
included in the revised catalog of bright {\em IRAS} galaxies
\citep{sand2003}, despite their close proximity and large mass.  Their
8--120~$\micron$ luminosities, based on {\em IRAS} fluxes from
\citet{mosh1990} for NGC~5746 and \citet{rice1988} for NGC~5170,
translate into global SFRs of $0.8\pm 0.2$ (NGC~5746) and $0.3\pm
0.1$~M$_{\sun}$~yr$^{-1}$ (NGC~5170), using the relations of
\citet{sand1996} and \citet{kenn1998}.  This is equivalent to specific
SFRs of $2.8\pm 0.7 \times 10^{-4}$ (NGC~5746) and $1.3\pm 0.4\times
10^{-4}$~M$_{\sun}$~yr$^{-1}$~kpc$^{-2}$ (NGC~5170), assuming a
circular disk of radius equal to the semi-major axis of $D_{25}$, the
ellipse outlining a $B$--band isophotal level of 25~mag~arcsec$^{-2}$.
This could be compared to the value of $\sim 2$ M$_{\sun}$ yr$^{-1}$,
i.e.\ $\sim 3\times 10^{-3}$~M$_{\sun}$~yr$^{-1}$~kpc$^{-2}$, for the
moderately quiescent Milky-Way disk (see e.g.\ \citealt{casu2004}).
From their {\em IRAS} 60 and 100~$\micron$ fluxes $S_{60}$ and
$S_{100}$, both galaxies furthermore display low far-infrared
'temperatures' of $S_{60}/S_{100}=0.13\pm 0.02$ (NGC~5746) and
$0.28\pm 0.05$ (NGC~5170), and 'mass-normalized' star formation rates
$L_{\rm FIR}/L_{\rm B}$ of $0.048\pm 0.012$ and $0.037\pm 0.012$.
These values are also indicative of low star formation activity (see
e.g.\ \citealt{read2001}).  Salient features of the observed galaxies
are listed in Table~\ref{tab,galdata}.

\begin{table*}
\begin{center}
\caption{Salient features of the observed disk galaxies}\label{tab,galdata}
\begin{tabular}{cccccccccc}
  \tableline\tableline
  Name & $D$ & RA& Dec& Hubble& $M_{\rm B}$& $D_{25}$& $i$ & $v_c$ &$N_{\rm H}$\\
  &(Mpc)&(J2000)&(J2000)&Type& &(arcmin)& &(km s$^{-1}$) 
  &(cm$^{-2}$)\\
  \tableline
 NGC~5746 & 29.4 & $14^h44^m56\fs 00$ & $+01\degr 57\arcmin17\farcs 1$ & SBb & $-21.79$  & (6.92, 1.20)&
  $84\degr$ & $318\pm10$ & $3.3\times 10^{20}$ \\
  NGC~5170 & 24.0 & $13^h29^m48\fs 83$ & $-17\degr 57\arcmin59\farcs 4$ & Sc & $-21.18$ & (8.32, 1.20) &
  $90\degr$ & $247\pm3$ & $7.3\times 10^{20}$ \\
  \tableline
\end{tabular}
\tablecomments{Distances taken from \citet{tull1988} (assuming
  $H_0=75$ km s$^{-1}$ Mpc$^{-1}$) and positions from the NASA/IPAC
  Extragalactic Database (NED).  Absolute $B$ magnitude, $D_{25}$
  (major and minor diameter), inclination $i$, and disk circular
  velocity $v_c$ were taken from the Hyperleda database.  Absorbing
  column density $N_{\rm H}$ is the Galactic value from
  \citet{star1992}.}
\end{center}
\end{table*}

\section{Observations and analysis}\label{sec,obs}

\subsection{Data Preparation}

Both disk galaxies were observed by {\it Chandra} (obs.\ ID 3928 and
3929) with the ACIS-I array as aimpoint and with the CCD's at a
temperature of $-120^\circ$~C. Data were telemetered in Very Faint
mode which allows for superior background suppression relative to
standard Faint mode.  To exploit this, the data were reprocessed and
background screened using {\sc ciao} v3.3.  Bad pixels were screened
out using the bad pixel map provided by the reduction pipeline, and
remaining events were grade filtered, excluding {\em ASCA} grades 1,
5, and 7.  Periods of high background on the ACIS-I chips were
filtered using $3\sigma$ clipping of the 0.3--12~keV lightcurves
extracted in off-source regions in 200~s bins. Resulting lightcurves
showed no flaring periods, leaving a total of 36.8~ks (NGC~5746) and
33.0~ks (NGC~5170) of cleaned exposure time.  Blank-sky background
data from the calibration database were screened and filtered as for
source data, and reprojected to match the aspect solution of the
latter. They were subsequently scaled to match our source data
according to the 10--12~keV count rates (these deviated by an
acceptable 5--6~per~cent from the exposure ratio between source and
background data).

An important issue for obtaining robust background estimates and
singling out any halo X-ray emission is the ability to efficiently
identify and remove point sources. This was a key limiting factor in
the similar {\em ROSAT} study by \citet{bens2000}.  The much narrower
point spread function of {\em Chandra} allows us to easily excise
point source regions without losing much detector area, thus
preserving most of the available source and background counts. We
identified point sources on the ACIS-I array using the wavelet-based
{\sc ciao} tool {\tt wavdetect}, adopting a threshold significance of
$1.8\times 10^{-6}$ for each CCD, which should limit the number of
spurious detections to $\sim 1$ per CCD.  A total of 126 (NGC~5746)
and 101 (NGC~5170) point sources were detected, in good agreement with
the statistical expectation of $\sim 25$ background sources per CCD
for a 35~ks pointing (cf.\ \citealt{summ2003}).  Inside $D_{25}$, we
detect 20 and 17 sources, respectively. All point sources were masked
out in the analysis of diffuse emission, using their $4\sigma$
detection ellipses from {\tt wavdetect}.

\subsection{X-ray and H$\alpha$ Imaging}

To aid the search for any diffuse X-ray emission in and around the
galactic disks, adaptively smoothed, background-subtracted, and
exposure-corrected {\em Chandra} images of the central region around
each galaxy were produced using a method similar to the one described
by \citet{rasm2006}. Source images were first smoothed using the {\sc
  asmooth} algorithm \citep{ebel2006} (based on the {\sc csmooth}
procedure in {\sc ciao}), which employs a kernel size that adapts
itself to achieve a certain signal-to-noise (S/N) level under the
kernel. We adopted a significance interval of 3--5$\sigma$, so that
the local smoothing scale is increased until an S/N ratio of 3 is
reached under the kernel, while features significant at more then
5$\sigma$ are left unsmoothed. Background maps were generated from
blank-sky data and scaled to match the source count rates in
source-free regions on the relevant chip. Using the map of adopted
smoothing scales returned by the smoothing algorithm, the background
images were then smoothed to the same spatial scales as the source
data and subtracted from the latter. The resulting images were finally
exposure-corrected using similarly smoothed, spectrally weighted
exposure maps, with weights derived from spectral fits to the
integrated diffuse emission. We stress that the resulting images
served illustration purposes only and were not used for any
quantitative analyses.

In addition to the X-ray data, we also obtained H$\alpha$ images of
both galaxies with the Danish 1.54-m telescope at La Silla, Chile, for
total exposures of 120 (NGC~5746) and 160~min (NGC~5170). In both
cases, 20~min $R$-band images were also taken for subsequent
continuum-subtraction.  The frames were bias-subtracted, flat-fielded,
and median-combined using standard {\sc iraf} procedures.  Using the
$R$-band image of each galaxy, we estimated and subtracted the
continuum emission contribution to the H$\alpha$ image as follows.
The two combined images were normalized to the same exposure time, and
the $R$-band image was smoothed with a Gaussian kernel to match the
broader point spread function of the H$\alpha$ image. The smoothed
$R$-band image was then scaled to match the narrower width of the
H$\alpha$ filter and subtracted from the H$\alpha$ image.  For
NGC~5746, the background in the resulting continuum-subtracted
H$\alpha$ image had a large-scale gradient caused by stray light from
a star just outside the field of view. We modeled this using the {\sc
  SExtractor} software \citep{bert1996} with options set to save a
full resolution interpolated background map. Parameters were optimized
to generate a map sufficiently fine to represent large scale
variations, and sufficiently coarse so as not to include local
non-background structures (such as an H$\alpha$ halo around the
galaxy). The resulting background map was then subtracted from the
H$\alpha$ image.

\subsection{X-ray Spectroscopy and Background
  Estimation}\label{sec,XRB}

Given that any hot gaseous halo could cover a large fraction of the
{\em Chandra} field-of-view in both cases, using our {\em Chandra}
data for background estimation ('local' background subtraction) may
not produce reliable results.  Instead, the appropriate blank-sky
background data from the calibration database were employed for
background estimates, thus also taking advantage of the superior
statistics of these background data.  However, due to spatial
variations in the soft X-ray background, the nominal {\em ROSAT}
All-Sky Survey 0.5--0.9~keV fluxes at our two pointings are a factor
$\sim 1.5$ (NGC~5746) and $\sim1.3$ (NGC~5170) higher than the
exposure-weighted mean value of the resulting blank-sky data.  In
order to account for this excess emission in spectral analysis, we
adopted the commonly used 'double--subtraction' technique (e.g.\
\citealt{arna2002}). We first extracted source spectra from which
blank-sky data of the same detector region were then subtracted. The
result includes any local soft X-ray background excess at our source
pointings relative to the blank-sky data. We then also extracted
spectra in off-source regions ($7\arcmin$--$8\arcmin$ annuli) from our
data and subtracted the associated blank-sky spectra, the result
reflecting the spectrum of this soft background excess at our source
pointings.  These residual spectra were finally also subtracted from
the blank-sky subtracted source-region spectra.  Spectrally weighted
response files were generated for the appropriate detector regions
using {\sc ciao}.

An implication of the results described in the following section is
that there could be residual source flux at the 20~per~cent level in
the $7\arcmin$--$8\arcmin$ annulus used for estimating the local
background excess relative to the blank-sky data in the spectral
analysis of NGC~5746. Since this local excess constitutes roughly half
of the estimated total 0.3--2~keV background for NGC~5746, we could be
underestimating the flux of the background-subtracted source spectrum
by about 10~per~cent in this band. However, vignetting differences
between the source region and this annulus almost exactly compensates
for this effect in terms of the total 0.3--2~keV flux, yielding a
soft-band source flux which should be accurate to within 5~per~cent in
the spectral analysis. This is comparable to the statistical
uncertainty on this quantity. Thus, though the distortions introduced
to the total background spectrum by these competing effects should be
small compared to the statistical uncertainty in each spectral bin, we
have added a 20~per~cent systematic error in quadrature to the
statistical errors on all parameters obtained from spectral fitting of
the NGC~5746 data.

For the spectral model fitting, a standard approach is to accumulate
the spectra in bins of at least 20~net~counts, and perform the fits
assuming $\chi^2$ statistics.  However, since the diffuse emission in
either galaxy is rather faint, we opted for a different approach in an
attempt to maximally exploit the limited spectral information
available. Instead, spectra were accumulated in bins of at least 5~net
counts and fitted using Cash statistics. Since the Cash statistic does
not provide a straightforward measure of the goodness of fit, we
tested the quality of each fit by evaluating the corresponding reduced
$\chi^2$ statistic for the data binned into 20~cts~bin$^{-1}$. This
allowed us to readily reject models that were not statistically
acceptable. As a second test, we confirmed that all fit results
obtained in this way were consistent with those of the ``standard''
method.  All fits were performed in the 0.3--5~keV band using {\sc
  xspec} v11.3, assuming the Solar abundance table of
\citet{ande1989}.

\section{Results}\label{sec,results}

\subsection{X-ray and H$\alpha$ Images}

Unprocessed {\em Chandra} images of each galaxy do not show any clear
evidence for diffuse emission outside the disk regions. Adaptively
smoothed 0.3--2~keV images, generated as described in
\S~\ref{sec,sample}, are shown in Figure~\ref{fig,overlays}, along
with the optical extent ($D_{25}$) of the galaxies (note that a
different version of the NGC~5746 image was shown in Paper~I, in which
each point source region prior to smoothing had been refilled by a
Poisson distribution interpolated from a surrounding region using the
{\sc ciao} task {\tt dmfilth}).  Although the resulting images should
be interpreted cautiously, particularly given the fairly low X-ray
surface brightness outside the disk of both galaxies, the smoothed
images do display evidence for diffuse emission in the disks of both
galaxies, with an indication that for NGC~5746 this emission extends
well beyond the disk midplane.  This is perhaps more easily seen in
the center panel of Figure~\ref{fig,overlays}, which shows our reduced
H$\alpha$ images of the two galaxies with the X-ray contours
overlayed.  Finally, in order to further enable a comparison to the
radio continuum morphologies of the galaxies, the right panel shows
again the H$\alpha$ data, this time with NRAO VLA Sky Survey 1.4~GHz
data overlayed \citep{cond1998}.  This panel also displays the
extraction regions discussed in the text below.

\begin{figure*}
\begin{center}
\epsscale{1.0}
\plotone{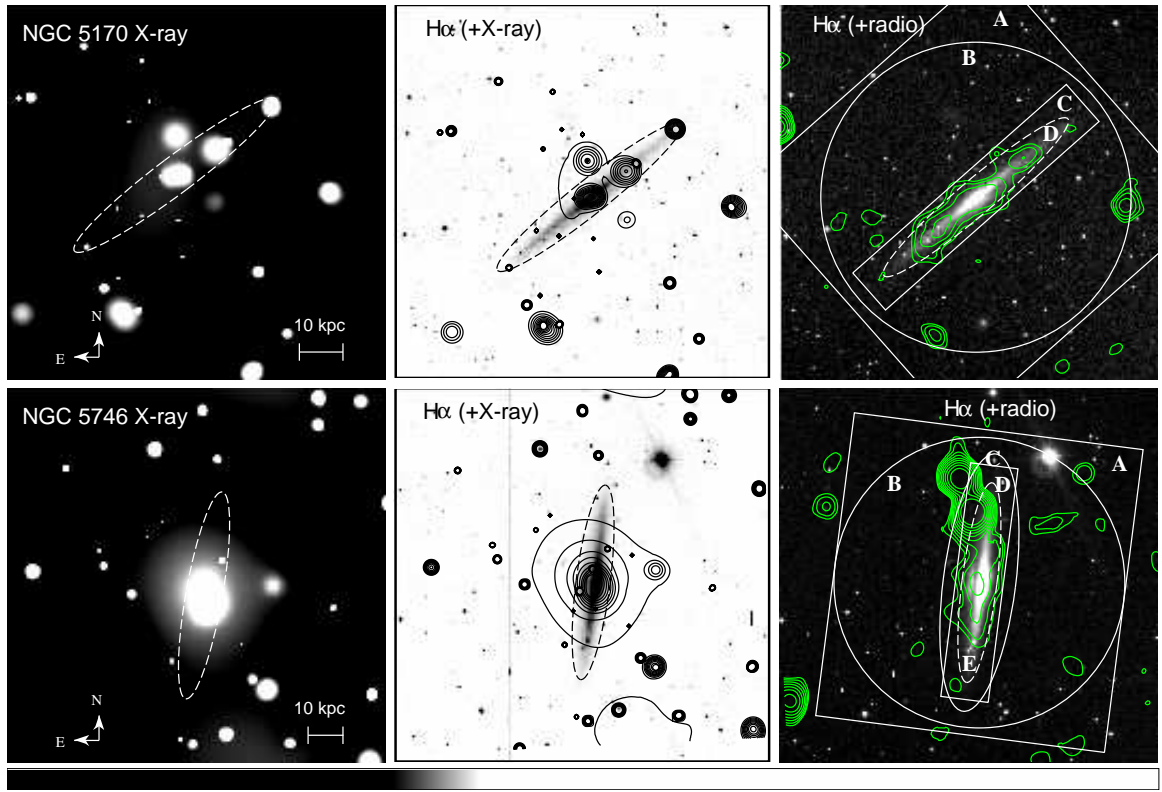}
\end{center}
\figcaption{Left: Adaptively smoothed 0.3--2.0~keV {\em Chandra}
  images of the central $13\times 13$~arcmin region around each
  galaxy. The images have been background-subtracted and
  exposure-corrected. Center: H$\alpha$ images with X-ray contours
  from the left figure overlayed.  Contours are logarithmically spaced
  over a decade, beginning at $6.5\times
  10^{-8}$~photons~cm$^{-2}$~s$^{-1}$~arcmin$^{-2}$ for both galaxies.
  Right: H$\alpha$ images with NVSS 1.4~GHz contours, beginning at
  1~mJy~beam$^{-1}$ and spaced by a factor of $\sqrt{2}$. Also shown are the
  regions used for extraction of X-ray surface brightness profiles
  ('A') and spectra (remaining regions).  The innermost elliptical
  region shown in all figures outlines the $D_{25}$ ellipse.
  \label{fig,overlays}}
\end{figure*}

As can be seen, there is no significant evidence for diffuse H$\alpha$
or radio continuum emission extending well outside the optical disk of
either galaxy.  The two radio sources seen outside the optical disk of
NGC~5746 are most probably background sources. They are not detected
in our X-ray data despite the proximity of the galaxy, nor at
near-infrared (2MASS) or optical wavelengths, and their 1.4~GHz fluxes
of 38.9 and 27.1~mJy are not included in the value listed for the
NGC~5746 disk by \citet{cond1998}.  Moreover, radio sources in
star-forming galaxies are known to form a tight radio--far-infrared
correlation, with a ratio of 60~$\micron$ to 1.4~GHz flux of
log\,$(S_{60}/S_{\rm 1.4~GHz})\approx 2.15$
\citep*{cond1986,cond1991}. Assuming the two sources to be associated
with NGC~5746, this relation would suggest their combined 60~$\micron$
output to exceed that of the entire galaxy ($1.3\pm 0.1$~Jy;
\citealt{mosh1990}) by a factor of $\sim 7$. This makes us conclude
that these two sources are most likely associated with radio-loud
background AGN.

A potential caveat associated with the interpretation of
Fig.~\ref{fig,overlays} is that the extraplanar emission apparent in
this Figure could represent disk emission which has been smoothed
beyond the disk by our adopted smoothing algorithm. Hence, in order to
perform a more quantitative search for diffuse X-ray emission in and
around the galaxies, and confirm the overall impression conveyed by
Fig.~\ref{fig,overlays}, we generated 0.3--2~keV surface brightness
profiles of the unsmoothed, exposure-corrected emission.  The profiles
were extracted perpendicular to the disks, within the rectangular
regions labelled 'A' in Fig.~\ref{fig,overlays}, and with point
sources masked out.  Results are shown in
Figure~\ref{fig,surfbright2}.  For NGC~5746, the individual surface
brightness profiles on either side of the disk are seen to be mutually
consistent within the errors.  The figure also shows the profiles of
both galaxies with the signal on both sides of the disk of each galaxy
co-added to improve statistics.  The corresponding profiles extracted
from the exposure-correced blank-sky background data within identical
apertures are shown as dotted error bars. These provide in indication
of the importance of CCD background variations and remaining
instrumental signatures within the regions examined, allowing a test
of whether the observed profiles can be explained by such effects.

\begin{figure*}
\begin{center}
\epsscale{1.0}
\plotone{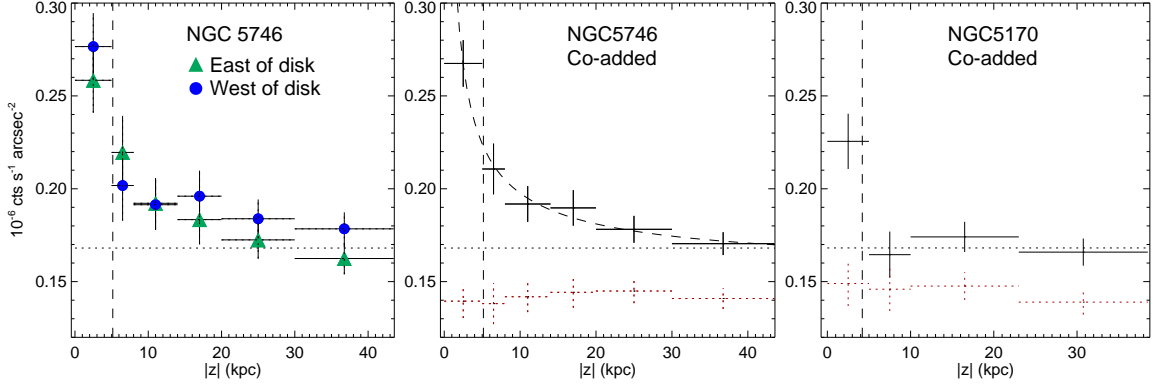}
\end{center}
\figcaption{Surface brightness profiles of diffuse X-ray emission. The
  extent of $D_{25}$ (minor axis) is marked by vertical dashed lines.
  Dotted data points show the level of the blank-sky background data
  prior to correction for excess soft X-ray background emission at our
  source pointings, while dotted horizontal lines represent the total
  mean background level estimated from 'double--subtraction'
  background spectra of the relevant regions, see
  \S~\ref{sec,XRB}. For the NGC~5746 co-added
  profile, the dashed curve represents the best-fitting power-law
  model.
\label{fig,surfbright2}}
\end{figure*}

Even if assuming no {\em a priori} knowledge of the background level
in our two {\em Chandra} observations, Fig.~\ref{fig,surfbright2}
gives a clear indication of the presence of net excess emission
outside the optical extent of NGC~5746. The result shown in
Fig.~\ref{fig,surfbright2} strongly suggests that the extended
extraplanar emission evident in Fig.~\ref{fig,overlays} is not an
artefact of the smoothing procedure adopted when generating the X-ray
image; excess emission is extending out to at least $\sim 20$~kpc on
either side of the disk, with surface brightness decreasing with
vertical distance to the disk. The profile bears no resemblance to
that of the background data, confirming that the observed signal is
real and not caused by instrumental artefacts or details of our
analysis.  For NGC~5170, on the other hand, the emission outside
$D_{25}$ is everywhere consistent with a constant level equal to the
estimated mean background level, and its distribution seems to follow
that of the blank-sky background. This result, obtained in an
identical way to that of NGC~5746, again confirms that the emission
seen around NGC~5746 is not a product of our analysis method, thus
supporting also the results indicated by the multi-wavelength
comparison in Fig.~\ref{fig,overlays}. The presence of excess emission
around NGC~5746 and the lack of any around NGC~5170 is further
corroborated by the results of spectral analyses, as described below.

We characterized the co-added surface brightness profile of NGC~5746
using a model comprising a power-law plus a spatially constant
background, $S = S_0 |z|^{-\Gamma} + B_0$.  We find $S_0 =
0.21_{-0.04}^{+0.07} \times 10^{-6}$~cts~s$^{-1}$ arcsec$^{-2}$,
$\Gamma= 0.68^{+0.42}_{-0.35}$, and $B_0=0.15_{-0.04}^{+0.02}$
cts~s$^{-1}$ arcsec$^{-2}$, with a reduced $\chi^2_\nu =0.67$ for
3~degrees of freedom $\nu$. The uncertainties on $\Gamma$ are large,
but we note that the best-fitting background level resulting from this
approach is entirely consistent with the value $B_0^{\prime} \approx
0.168\times 10^{-6}$~cts~s$^{-1}$ arcsec$^{-2}$ determined
independently from our spectral 'double-subtraction' background
analysis, shown as a dotted line in Fig.~\ref{fig,surfbright2}. If
fixing $B_0$ at this value in the fit to the profile, $\Gamma$ becomes
significantly better constrained to $\Gamma = 1.0\pm 0.1$, with $S_0 =
0.25^{+0.06}_{-0.05}\times 10^{-6}$~cts~s$^{-1}$ arcsec$^{-2}$. This
fit result is shown as a dashed curve in Fig.~\ref{fig,surfbright2}.

\subsection{X-ray Spectra}\label{sec,spectra}

For the extraction of spectra of the halo regions, the choice of
aperture inner limit was a compromise between sampling regions close
to the disk, where halo emission is expected to be strongly
concentrated \citep{toft2002}, and a need to remain insensitive to the
presence of any outflowing gas from the disk resulting from, for
example, star formation activity. However, given the low specific star
formation rates of the disks (of order
$10^{-4}$~M$_{\sun}$~yr$^{-1}$~kpc$^{-2}$, cf.\ \S~\ref{sec,sample})
one could {\em a priori} expect very little outflowing hot gas
resulting from such activity. The investigation presented in
\S~\ref{sec,origin} below indicates that this assertion is justified.
For both galaxies, many of the results presented below therefore apply
to the region immediately outside $D_{25}$. Spectra obtained at larger
distances from the disk produce consistent results but naturally
return larger uncertainties on fitted parameters.  For NGC~5746, the
aperture outer limit was chosen with the guidance of
Fig.~\ref{fig,surfbright2}, and with the general aim of maximizing the
S/N ratio.  For NGC~5170, with its smaller $v_c$ and hence presumably
lower halo $L_{\rm X}$, we simply used regions whose physical extent
are slightly smaller than those of NGC~5746.

Selected results for the different source regions labelled 'B'--'E' in
Fig.~\ref{fig,overlays} are presented in Table~\ref{tab,spec}.  For
NGC~5746, the background-subtracted spectrum underlying the bold-faced
halo result listed in the Table is shown in Figure~\ref{fig,spec}$a$.
Excess 0.3--2~keV emission outside $D_{25}$ is detected at $5\sigma$.
Its properties are well described by an optically thin thermal plasma
({\em mekal} model in {\sc xspec}), with $T=0.56^{+0.13}_{-0.18}$~keV.
From the fractional $1\sigma$ errors on the normalization of this
best-fitting spectral model, we derive a total unabsorbed 0.3--2~keV
halo luminosity of NGC~5746 of $7.3\pm 3.9 \times
10^{39}$~erg~s$^{-1}$.  The derived halo metallicity is low,
$Z=0.04^{+0.09}_{-0.02}$, but since statistics are too poor to
constrain any temperature variations within the halo, we cannot
exclude the possibility that the derived metallicity could be biased
downwards due to the well-known Fe bias (arising when fitting a
multi-temperature plasma with a single-temperature model; see
\citealt{buot2000}). For the derived temperature, the halo luminosity
remains within the quoted errors for all subsolar metallicities,
however.  We note that excess emission at $>4\sigma$ significance is
found well outside the disk in our spectral analysis (e.g., region
$B\setminus C$ in Table~\ref{tab,spec}). This confirms, independently
of the extracted surface brightness profile, that the extraplanar
emission seen in Fig.~\ref{fig,overlays} does not represent smoothed
disk emission.  For comparison, Figure~\ref{fig,spec}$b$ also shows a
background-subtracted spectrum of the NGC~5746 {\em disk} emission.
This is somewhat different from the halo spectrum, showing a tail at
higher energies, possibly due to the presence of unresolved point
sources in the disk.

\begin{table*}
\begin{scriptsize}
\begin{center}
  \caption{Results of spectral fitting to the diffuse X-ray emission\label{tab,spec}}
\begin{tabular}{ccccllc}
  \tableline\tableline
  Region & {\em H}alo/ & Net Cts & S/N & {\sc xspec} Model & 
  Fit Results & $\chi^2/\nu$ \\
  & {\em D}isk? & (0.3--2 keV) & (0.3--2 keV) &  &  \\
  \tableline 
  {\em NGC~5746} & &  &  &  &  &  \\
  {\bf B$\setminus$E} & {\bf {\em H}} & {\bf 230}  & {\bf 5.1} & {\bf {\em wabs}({\em mekal})} & $\mathbf{T=0.56_{-0.18}^{+0.13}}$, $\mathbf{Z=0.04^{+0.09}_{-0.02}}$~$\mathbf{Z_\odot}$ & {\bf 9.77/9} \\
  &     &      &     & {\em wabs}({\em pow})   & $\Gamma =2.98^{+1.72}_{-1.75}$ & 10.99/9 \\  
  B$\setminus$D  & {\em H} & 192  & 4.8 & {\em wabs}({\em mekal}) & $T=0.53_{-0.25}^{+0.45}$ & 5.53/8 \\
  B$\setminus$C  & {\em H} & 175  & 4.4 & {\em wabs}({\em mekal}) & $T=0.43_{-0.24}^{+0.68}$  & 5.83/7 \\
  E  & {\em D} & 122  & 10.6& {\em wabs}({\em pow+mekal}) & $\Gamma =0.44^{+0.40}_{-0.50}$, $T=0.36^{+0.27}_{-0.18}$ & 4.28/6 \\
  (=$D_{25}$)   &   &   &   & {\em wabs}({\em pow+2$\times$mekal}) & $\Gamma =0.43^{+0.42}_{-0.51}$, $T_1=0.13_{-?}^{+25.5}$, & 3.81/4 \\
  &  &   &   &                         &  $T_2= 0.71_{-0.30}^{+6.84}$ & \\
\tableline
 {\em NGC~5170} & &  &  &  & & \\ 
  {\bf B$\setminus$D}  & {\bf {\em H}} & {\bf 4} & {\bf 0.0} & {\bf \ldots} & {\bf \ldots} & {\bf \ldots} \\
  C  & {\em D}  & 71   & 4.4 & {\em wabs}({\em pow}) & $\Gamma = 1.83^{+1.46}_{-1.00}$ & 3.75/4 \\
  &     &      &     & {\em wabs}({\em mekal}) & $T=8.75_{-7.44}^{+?}$ & 4.03/4 \\
  D  & {\em D} & 69   & 4.9 & {\em wabs}({\em pow}) & $\Gamma = 0.72_{-0.57}^{+0.65}$ & 2.87/3 \\
  (=$D_{25}$)   &      &      &     & {\em wabs}({\em pow+mekal}) & $\Gamma =0.21^{+0.63}_{-0.70}$, $T=0.08_{-?}^{+1.94}$  & 0.28/1 \\
  \tableline
\end{tabular}
\tablecomments{Source regions are depicted in Fig.~\ref{fig,overlays}
  ('B$\setminus$D' means 'B excluding D'). Spectral model components
  are absorbed power-laws (intensity $\propto E^{-\Gamma}$) and
  thermal plasmas (``{\em mekal}''). The Galactic value of $N_{\rm H}$
  (Table~\ref{tab,galdata}) is assumed for the absorbing component,
  with $Z=0.1$~Z$_{\sun}$ and $Z=1.0$~Z$_{\sun}$ for the halo and disk
  metallicities, respectively, where not specified otherwise.  All
  temperatures are in keV. A '?' means that the uncertainty is
  unconstrained. Last column gives the goodness of fit for the number
  of degrees of freedom $\nu$. Fit results for which $\chi^2_\nu > 2$
  are not presented.}
\end{center}
\end{scriptsize}
\end{table*}

\begin{figure*}
\begin{center}
\epsscale{1.0}
\plotone{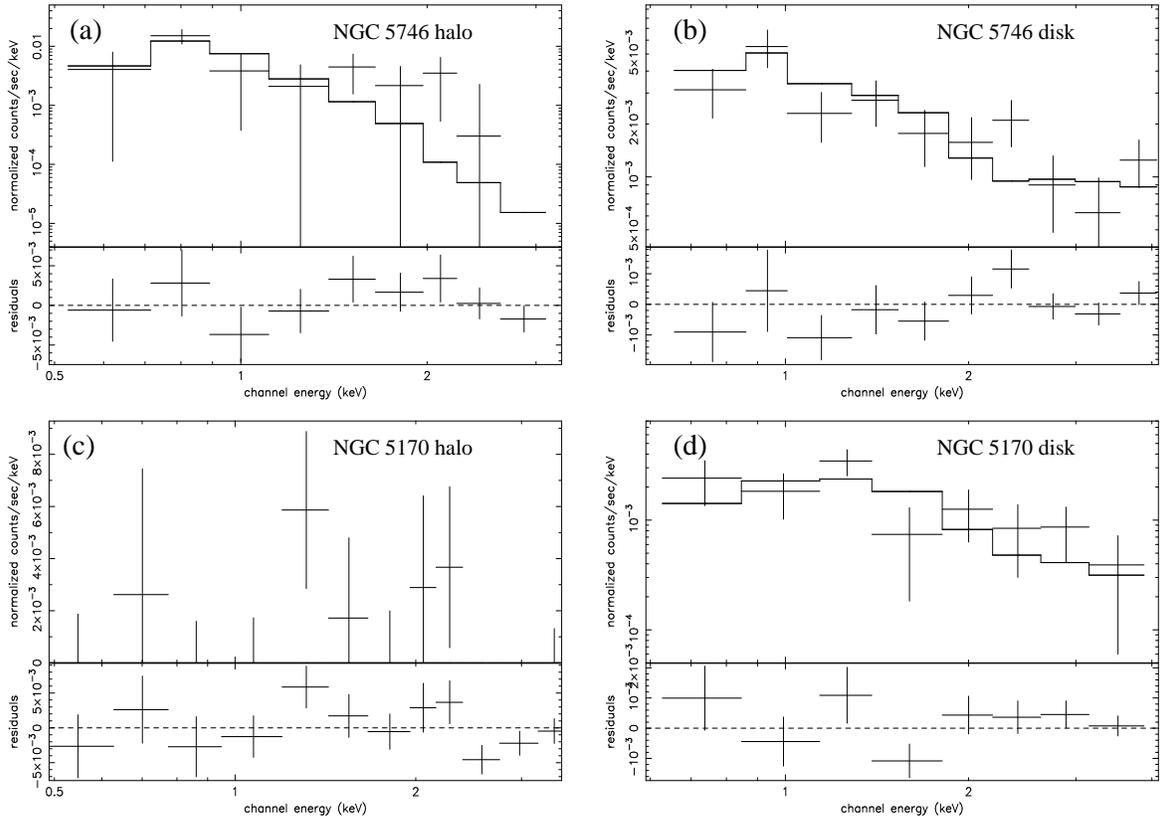}
\figcaption{(a) Background-subtracted spectrum of NGC~5746 diffuse
  halo emission (region B$\setminus$E of Table~\ref{tab,spec}) along
  with the best-fitting thermal plasma model.  Bottom panel shows fit
  residuals. (b) Corresponding plot for the NGC~5746 diffuse disk
  emission (region E of Table~\ref{tab,spec}) along with the
  best-fitting thermal plasma + power-law model. (c) Spectrum of the
  NGC~5170 halo region (region B$\setminus$D of Table~\ref{tab,spec}).
  Residuals show the deviation from zero net emission. (d) As (b), for
  the NGC~5170 disk emission (region D of Table~\ref{tab,spec}).
\label{fig,spec}}
\end{center}
\end{figure*}

For NGC~5170, no net emission is detected outside $D_{25}$, where the
derived emission level remains consistent (at $1\sigma$) with that of
the background, as can be seen in Fig.~\ref{fig,spec}$c$. An
assumption about the halo temperature of this galaxy must therefore be
made in order to constrain its $L_{\rm X}$.  To this end, we note that
simulations predict a remarkably tight correlation between the
temperature of infalling hot gas and $v_c$ of the disk
\citep{toft2002}, with NGC~5746 falling onto this trend (see
\S~\ref{sec,comparison}).  Hence, if assuming $T \approx 0.2$ keV for
the hot halo of NGC~5170, as suggested by its disk circular velocity
of $v_c \approx 250$~km~s$^{-1}$ \citep{toft2002} or by a simple
expectation for the virial temperature $T_{\rm vir}\approx (1/2)\mu
m_p v_c^2$ of the underlying dark matter halo, we obtain an upper
limit to its 0.3--2~keV luminosity of $L_{\rm X} < 1.2\times
10^{39}$~erg~s$^{-1}$ for any subsolar metallicity. This upper limit
is relaxed to $\sim 1.7\times 10^{39}$~erg~s$^{-1}$ if instead
assuming, for example, $T\approx 0.15$~keV, as would be appropriate
for the significantly less massive Milky Way.

The disk emission of NGC~5746 cannot be described by a thermal
component alone ($\chi^2_\nu = 4.38$ for 8 d.o.f.) but is well
described by a thermal plasma of temperature $T\approx 0.3$~keV plus a
power-law component, a result very similar to those obtained for the
disks of other spirals, both quiescent galaxies and starbursts (e.g.\
\citealt{read1997,rasm2004b} and references therein).  The thermal
emission is most likely a signature of hot gas, shock heated by
supernova explosions, whereas the power-law component may represent
the integrated emission from a population of unresolved point sources,
presumably mainly X-ray binaries and supernova remnants.  Corrected
for Galactic absorption and assuming $Z=$~Z$_\odot$, the total
0.3--2~keV luminosity of the NGC~5746 emission inside $D_{25}$ is
$\approx 3.7\times 10^{39}$~erg~s$^{-1}$, of which
$1.2^{+0.6}_{-0.4}\times 10^{39}$~erg~s$^{-1}$, i.e.\ $\sim 30\pm
10$~per~cent, is due to the power-law component.  Some spiral disks
show spectral evidence for a second thermal component, but while our
data do not rule out this possibility for the NGC~5746 disk,
statistics are too poor to obtain useful constraints on such a model
from spectral fitting (see Table~\ref{tab,spec}).

For the same reason, the properties of the disk emission of NGC~5170
are not robustly characterized by our data.  A thermal model fit to
the disk spectrum of this galaxy, shown in Fig.~\ref{fig,spec}$d$,
produces a physically absurd temperature of $T>80$~keV, whereas a
simple power-law provides an adequate description of the data.  The
reason could be strong intrinsic absorption in the disk, hardening the
X-ray spectrum beyond that reproducible by a simple thermal model.  A
total unabsorbed 0.3--2~keV disk luminosity of
$0.8_{-0.3}^{+0.4}\times 10^{39}$~erg~s$^{-1}$ is derived for this
galaxy.

\section{Properties of the X-ray gas and comparison to other X-ray halos}
\label{sec,prop}

As we have seen, there is evidence from both imaging and spectra of
diffuse X-ray emission surrounding NGC~5746 out to at least $|z|\sim
20$~kpc from the disk midplane.  NGC~5170, on the other hand, does not
display evidence of any detectable extraplanar X-ray emission.  It is
worth emphasizing that this notable difference between the two
galaxies emerges after having subjected both data sets to the exact
same data reduction and analysis procedure. This strongly suggests
that the apparent surface brightness and spatial behaviour of the
NGC~5746 extraplanar emission are genuine features, which cannot
simply be ascribed to instrument characteristics, inaccurate
background subtraction, or other details of our analysis method.
Though a $5\sigma$ detection must be viewed as tentative rather than
conclusive, we regard it as a sufficiently strong result to warrant a
detailed discussion.  The implications of a non-detection
around NGC~5170 are also worth considering.

\subsection{Properties of Halo and Disk Gas}\label{sec,halogas}

Before discussing the properties of the detected emission in detail,
we will briefly review whether the halo emission around NGC~5746 is
truly diffuse or, given its low surface brightness, could potentially
originate in a population of unresolved point sources.  To test this,
we repeated our source detection procedure using lower thresholds to
see if sources just below the adopted threshold were present, but this
was not found to be the case.  Given the distribution of emission as a
function of distance from the NGC~5746 disk
(Fig.~\ref{fig,surfbright2}), any unresolved sources would have to be
associated with NGC~5746 and so must on average have $L_{\rm X}<
10^{38}$ erg s$^{-1}$ if assuming the halo spectrum of
Fig.~\ref{fig,spec}$a$. Although this is consistent with a population
of X-ray binaries, the low emission level above $E\approx 2$ keV and
the associated value of $\Gamma \sim 3$ for the best-fitting power-law
do not fit well into this scenario. A significant contribution from
either high-mass X-ray binaries or core-collapse supernova remnants is
furthermore ruled out by the low galactic star formation rate. If the
observed X-ray signal is, in turn, produced by low-mass X-ray
binaries, these being tracers of stellar mass at an estimated level of
$10^{10}$~M$_{\sun}$ of stars per $\sim 8\times 10^{38}$~erg~s$^{-1}$
\citep{gilf2004}, the derived value of halo $L_{\rm X}$ would suggest
the presence of several $10^{10}$ M$_{\sun}$ of stars outside the disk
region, another improbable scenario.  This picture is made less
probable still by the fact that a thermal plasma provides a marginally
better spectral fit to the emission than a power-law model.  In
summary, the emission must be regarded as truly diffuse and is most
probably due to a thermal plasma residing in an extended halo
surrounding the galaxy.

Using the parameters derived for the NGC~5746 halo emission from the
spectral analysis, the physical properties of the X-ray emitting halo
gas can be assessed.  From the spectral normalization $A$ of the
best-fit {\em mekal} model in {\sc xspec},
\begin{equation}\label{eq,xspec}
  A = \frac{10^{-14}}{4\pi D^2} \int{n_e n_H \mbox{ d}V},
\end{equation}
where $D$ is the distance (Table~\ref{tab,galdata}), one can obtain
the emission integral $EI \equiv \int{n_e n_H \mbox{ d}V} \approx \eta
n_e^2 V$ of the hot gas. Here $\eta$ is the volume filling factor of
the gas inside a total volume $V$; we note that simulations
(\citealt{toft2002}; \S~\ref{sec,comparison}) indicate high values of
$\eta \approx$ 0.8--1, and that $EI$ furthermore scales as $Z^{-1}$,
where $Z$ is the gas metallicity in solar units.  We assume the large
majority of detected X-ray gas to be confined inside a cylinder of
height $2\times 20$~kpc along the $D_{25}$ minor axis (cf.\
Fig.~\ref{fig,surfbright2}) and a base radius of 45 kpc
(Fig.~\ref{fig,overlays}).  The volume of the disk is excluded,
assuming this to be an ellipsoid of revolution with two identical
major axes and a minor axis both equal to those of the
$D_{25}$~ellipse.  This leaves a total volume $V= 6.9\times
10^{69}$~cm$^3$, which can be combined with the results for $EI$ and
the halo temperature $kT = 0.56^{+0.13}_{-0.18}$~keV, to derive
various properties of the hot gas.  Although these results will only
be estimates and averages within the cylinder, we include here for
completeness the formal statistical errors derived by propagating the
errors on $EI$, $T$, and $Z$.  For $\eta =1$ we find a mean electron
density $\langle n_e \rangle \sim (EI/V\eta)^{1/2} \approx
5.2^{+1.2}_{-1.7}\times 10^{-4}$~cm$^{-3}$, a gas mass $M_{\rm gas}
\sim m_p \langle n_e \rangle V \eta^{1/2} \approx 3.3\pm 1.0 \times
10^9$~M$_{\sun}$, a mean cooling time $\langle t_{\rm cool} \rangle
\sim 3kT\eta^{1/2}/(\Lambda \langle n_e \rangle) \approx
1.4^{+0.7}_{-0.3} \times 10^{10}$~yr, and a corresponding gas cool-out
rate $\langle \dot M \rangle \sim M_{\rm gas}/\langle t_{\rm
  cool}\rangle \approx 0.24\pm 0.13$~M$_{\sun}$~yr$^{-1}$.  Here we
have used the cooling curves $\Lambda(T,Z)$ of \citet{suth1993}.

In the same approximation, the bulk properties of any halo gas around
NGC~5170 can be constrained using similar arguments.  Given the lack
of halo detection, we must rely on simulation results for the gas
temperature, and we take $T\approx 0.2$~keV as described earlier.
Inside a volume derived the same way as for NGC~5746, and assuming
$Z=0.1$~Z$_{\sun}$, the derived upper limit to $L_{\rm X}$ can then be
translated into corresponding limits to the gas parameters of $\langle
n_e \rangle < 2.2\times 10^{-4}$~cm$^{-3}$, $M_{\rm gas} < 1.2\times
10^9$~M$_{\sun}$, $\langle t_{\rm cool} \rangle > 6.9\times 10^9$~yr,
and $\langle \dot M \rangle < 0.2$~M$_{\sun}$~yr$^{-1}$.
Table~\ref{tab,halos} summarizes the derived constraints on halo
properties for both galaxies.

\begin{table*}
\begin{center}
\caption{Derived constraints on hot halo gas\label{tab,halos}}
\begin{tabular}{ccccccccc}
  \tableline\tableline
  Name & $r$ & $T$ & $Z$ & $L_{\rm X}$ & $\langle n_e \rangle$ & $M_{\rm gas}$ & $\langle t_{\rm cool}\rangle$ & $\langle \dot M \rangle $ \\
  &(kpc)&(keV)&(Z$_\odot$)&(erg~s$^{-1}$)& ($cm^{-3}$) & (M$_\odot$) & (Gyr) & M$_\odot$~yr$^{1}$ \\
  \tableline
  NGC~5746 & 43 & $0.56^{+0.13}_{-0.18}$ & $0.04^{+0.09}_{-0.02}$ & 
  $7.3\pm 3.9 \times 10^{39}$ & $5.2^{+1.2}_{-1.7}\times 10^{-4}$ & 
  $3.3\pm 1.0\times 10^9$ & $14^{+7}_{-3}$ & $0.24\pm 0.13$ \\
  NGC~5170 & 35 & 0.2 (fixed) & 0.1 (fixed) & $<1.2\times 10^{39}$ & 
  $<2.2\times 10^{-4}$ & $<1.2\times 10^9$ & $>6.9$ & $<0.2$ \\
  \tableline
\end{tabular}
\tablecomments{$r$ is the outer radius of the aperture used for the
  extraction of mean halo properties. Luminosities are in the
  0.3--2~keV band.}
\end{center}
\end{table*}

For the X-ray gas in the disk of either galaxy, the filling factor and
the intrinsic absorption remain largely unknown as they cannot easily
be constrained from the present data. We will therefore refrain from
attempting to constrain other quantities than the spectral properties
of the disk gas already discussed in \S~\ref{sec,spectra}.  With the
disks having far-infrared 40--120~$\micron$ luminosities of $L_{\rm
  FIR}=1.4\times 10^{43}$ (NGC~5746) and $5.6\times
10^{42}$~erg~s$^{-1}$ (NGC~5170), we note that their estimated diffuse
X-ray luminosities render them consistent with the nominal $L_{\rm
  X}$--$L_{\rm FIR}$ relation derived by \citet{read2001} for the hot
gas in 'normal' (i.e.\ non-starburst) spirals, log $L_{\rm X} \approx
1.1$~log $L_{\rm FIR} -7.9$.  As $L_{\rm FIR}$ is known to be a tracer
of the star formation rate (e.g.\ \citealt{sand1996}), these results
provide further testimony to the modest star formation activity of
both galaxies.

\subsection{Comparison to Other Galactic X-ray Halos}\label{sec,halos}

In other cases where diffuse X-ray emission has been reported around
spirals, this emission is typically accompanied by extended H$\alpha$
and radio halos, with a good correspondence between all three
wavebands in terms of the extent and spatial structure of the
extraplanar emission (e.g., \citealt{stri2004, tull2006a}).  In
particular, all existing studies that have reported the detection of
extraplanar X-ray gas in 'normal' (moderately star-forming) spirals
also present clear evidence of coincident extraplanar H$\alpha$
emission. To our knowledge, the only exception to this is NGC~2613
\citep{li2006}, for which H$\alpha$ data do not appear to have been
reported in the literature.

In these cases, the origin of these multi-phase halos have been linked
to star formation activity in the disk
\citep{wang2001,wang2003,ehle2004,stri2004,tull2006a}.  It is well
established that significant star formation activity in spirals is
typically accompanied by the presence of diffuse ionized gas above the
disk midplane, with temperatures of a few times $10^4$ K (e.g.\
\citealt{ross2004}). The standard interpretation of the extraplanar
X-ray, H$\alpha$, and radio emission seen around other spirals is that
the emission is indeed associated with multiple phases of gas, all
expelled from the disk: The winds of supernovae and massive stars lift
hot, X-ray emitting gas above the plane, along with warm and/or
photoionized H$\alpha$-emitting material, with the radio halo produced
through synchrotron emission by SN-generated cosmic ray electrons
which couple to the magnetic field of the outflowing gas (e.g.\
\citealt{wang2001,wang2003}).

Although there is no indication in Fig.~\ref{fig,overlays} of
significant extraplanar H$\alpha$ or radio emission around NGC~5746,
we investigate this possibility in more detail in
Fig.~\ref{fig,Hasurf}, where we show the surface brightness profile of
the H$\alpha$ emission perpendicular to the disk.  As can be seen,
there is no evidence for significant amounts of H$\alpha$ gas outside
the optical extent of the galaxy.  Note that the dip in the H$\alpha$
profile 10~arcsec east of the $D_{25}$ center is due to the dust lane
also visible in Fig.~\ref{fig,overlays}.  NGC~5170 has been studied in
a similar way by \citet{ross2000} who find no evidence for extraplanar
diffuse H$\alpha$ emission surrounding this galaxy in an H$\alpha$
exposure of similar depth to ours.  There is also no indication from
Fig.~\ref{fig,overlays} of significant extraplanar {\em radio}
emission around either galaxy from VLA 1.4~GHz data, a conclusion
supported for NGC~5746 by the VLA flux density profile also displayed
in Fig.~\ref{fig,Hasurf}.

\begin{figure*}
\begin{center}
\epsscale{1.0}
\plotone{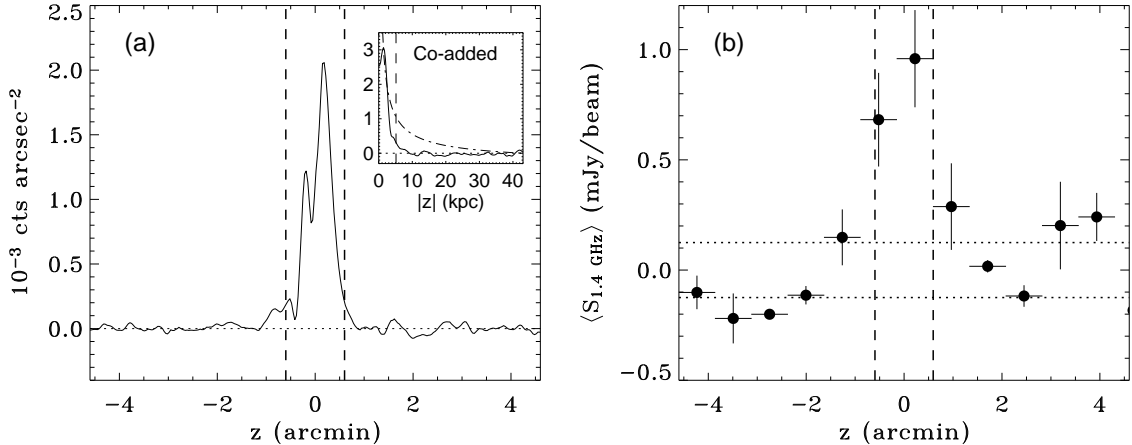}
\end{center}
\figcaption{(a) Continuum-subtracted H$\alpha$ surface brightness
  profile of NGC~5746, extracted within the same aperture as that
  underlying Fig.~\ref{fig,surfbright2}. Bright stars have been masked
  out.  Dashed vertical lines outline $D_{25}$ as in
  Fig.~\ref{fig,surfbright2}, with the East side of the disk to the
  left. The inset shows the co-added H$\alpha$ profile (same units).
  The dot-dashed curve marks the best-fit power-law to the X-ray
  surface brightness, scaled to match the peak of the H$\alpha$
  profile. (b) Corresponding mean flux density of 1.4~GHz continuum
  emission within the same aperture, in bins corresponding to the VLA
  beam size of 45~arcsec (FWHM). Bright point sources have been masked
  out. Dotted lines outline the $1\sigma$ errors on the noise level.
\label{fig,Hasurf}}
\end{figure*}

The absence of detectable extraplanar H$\alpha$ and radio emission in
NGC~5746 is thus unique among spirals with X-ray halo emission.  This
suggests that the observed X-ray emission around NGC~5746 may have a
different origin than in the standard case where the emission can be
linked to supernova-driven outflows. Another indication that this is
the case can be based on comparing the X-ray luminosity of the
extraplanar gas with the total ($8-120$~\micron) infrared luminosity
$L_{\rm IR}$ of the disk, adopting the standard assumption that the
latter is a measure of the global disk star formation rate (see, e.g.,
\citealt{stri2004}).  If the observed extraplanar X-ray emission
around NGC~5746 were due to starburst winds driven by core-collapse
SN, the ratio $L_{\rm X}/L_{\rm IR}$ should then reflect the
efficiency with which the mechanical energy output of SN in the disk
is converted into thermal energy of extraplanar X-ray gas.

A comparison between extraplanar X-ray emission around other spirals,
where detected, shows the ratio $L_{\rm X}/L_{\rm IR}$ to be confined
within a fairly narrow range, centered around a value which is in
reasonable agreement with theoretical expectations for SN outflows
\citep{stri2004}.  The situation is illustrated in
Figure~\ref{fig,strickland}, where we plot the infrared and
extraplanar 0.3--2~keV luminosities for all disk galaxies for which --
to our knowledge -- X-ray halo emission has been reported.  Included
are the galaxies in the sample of \citet{stri2004}, the additional
galaxies of \citet{tull2006a}, the recent result for NGC~2613 derived
by \citet{li2006}, as well as those galaxies of \citet{wang2004} that
are not included in the above samples.  Where necessary, we have
derived $L_{\rm IR}$ for these galaxies similarly to \citet{stri2004},
using {\em IRAS} fluxes from \citet{sand2003} where available, and
from \citet{mosh1990} otherwise.

\begin{figure}
\begin{center}
\epsscale{1.2}
\plotone{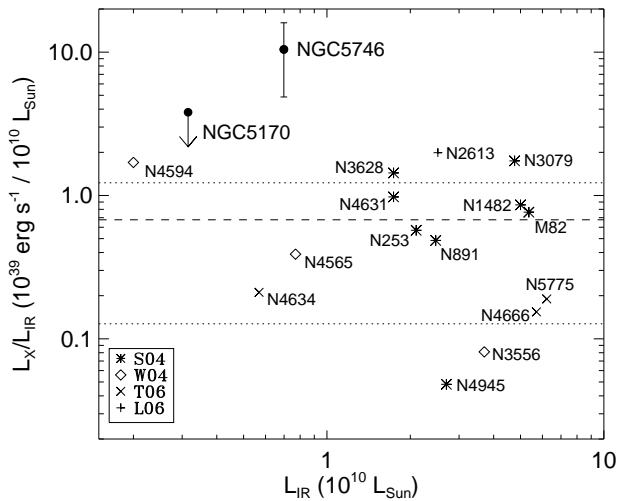}
\end{center}
\figcaption{Infrared and extraplanar X-ray luminosities for NGC~5746,
  NGC~5170, and for the galaxies with detected extraplanar X-ray
  emission in the sample of \citet[``S04'']{stri2004}. Also shown are
  the corresponding results for the additional galaxies of
  \citet[``T06'']{tull2006a}, \citet[``W04'']{wang2004}, and
  \citet[``L06'']{li2006}.  Dashed and dotted lines show the mean and
  $1\sigma$ dispersion for the S04 + T06 galaxies.
\label{fig,strickland}}
\end{figure}

In contrast to these samples, as can be seen in the Figure, the value
of $L_{\rm X}/L_{\rm IR}$ derived for NGC~5746 is unusually high.  The
mean and $1\sigma$ dispersion of $L_{\rm X}/L_{\rm IR}$ for the
galaxies of \citet{stri2004} and \citet{tull2006a} is $0.68\pm 0.55$
in the units displayed in Fig.~\ref{fig,strickland}, with the nominal
value for NGC~5746 lying $13.6\sigma$ above this mean. It is worth
pointing out that this difference arises, not because the halo $L_{\rm
  X}$ of NGC~5746 is conspicuously high, but because, as can be seen,
its value of $L_{\rm IR}$ is unusually low for the derived extraplanar
$L_{\rm X}$.  The immediate conclusion is that the NGC~5746
extraplanar emission seems to be of a different nature to that of the
galaxies in the comparison sample.

How does the extraplanar emission of NGC~5746 compare to that seen in
and around early-type galaxies? We note that the origin and heating
mechanism of X-ray gas in elliptical galaxies remain controversial
(e.g.\ \citealt{osul2004}), and it is possible that different
mechanisms are prevalent in low- and high-mass systems.  Entering that
debate is beyond the scope of this paper. A comparison is nonetheless
warranted on the basis that the X-ray luminosity of diffuse gas in
ellipticals roughly obeys $L_{\rm X}\propto L_B^2 \propto
\sigma^{6-7}$ (based on the Faber-Jackson relation and the $L_{\rm
  X}$--$L_B$ relation of \citealt{osul2001}), hence showing a similar
dependence on the characteristic stellar velocity of the system to the
one expected for the hot halos of spirals (\S~\ref{sec,intro}).  If
similar processes are responsible for the emission around NGC~5746,
one could perhaps expect this emission to follow similar scaling
relations as those obeyed by E/S0's.

The $L_{\rm X}$--$L_B$ and $L_{\rm X}$--$T$ relations derived for
X-ray bright ellipticals by \citet{osul2003} would nominally suggest
$L_{\rm X} \approx 2\times 10^{41}$ and $1\times
10^{41}$~erg~s$^{-1}$, respectively, given $L_B$ and (halo) $T$ for
NGC~5746. While these expectations lie well above the derived value
for the NGC~5746 halo, both relations are subject to substantial
scatter and can easily accommodate the NGC~5746 value.  In addition,
the agreement would improve if adding in the diffuse luminosity of the
NGC~5746 disk, which would probably provide a more fair comparison
(especially if also correcting this value for intrinsic absorption).

A comparison can also be made to the X-ray surface brightness profiles
of ellipticals, which are typically found to be well described by
standard $\beta$--models, with a mean value of $\beta \approx 0.55$
for X-ray luminous E's \citep{osul2003}, but again the scatter is
substantial.  For NGC~5746, the derived profile would correspond to
$\beta \approx 0.35$, which would place it at the very low end of the
range covered by these systems.  On the other hand, it is clear that
the profile derived for NGC~5746 must steepen beyond our radius of
detection, as the total halo luminosity would otherwise diverge (this,
of course, also applies to most of the ellipticals).

In summary, comparison to the diffuse emission of ellipticals is
somewhat inconclusive and to a certain degree hampered by the fact
that we are dealing with an edge-on galaxy, where (probably spatially
varying) intrinsic absorption in the disk precludes a robust estimate
of the {\em total} diffuse $L_{\rm X}$. Taken at face value, the halo
of NGC~5746 is less X-ray luminous for its disk $L_B$ and halo $T$
than the average X-ray bright elliptical, and also has a shallower
surface brightness profile.  However, the large scatter in the X-ray
properties of ellipticals precludes a firm statement as to whether the
diffuse X-ray emission around NGC~5746 is distinctively different from
that seen in early-type galaxies.

\section{Origin of NGC~5746 Extraplanar Emission}\label{sec,origin}

In Paper~I it was briefly substantiated that the excess X-ray emission
around NGC~5746 is most likely due to inflowing hot gas. Here we
consider in detail possible alternatives, demonstrating that a hot
gaseous halo of externally accreted gas indeed seems the most
plausible explanation for the origin of the detected extraplanar
emission.

\subsection{Gas Outflows from Stellar Activity}\label{sec,stars}

Given that NGC~5746 is isolated and hence should not be experiencing
ram-pressure or tidal stripping of its hot disk gas, the most
immediate explanation for its extraplanar X-ray emission would be
outflows of hot gas powered by stellar activity in the disk.  As
already discussed, such outflows could be driven by core-collapse
supernovae and mass loss from massive stars, or perhaps by SN~Ia in
the form of a bulge wind as suggested by \citet{wang2004}.  Although
the limited X-ray signal does not allow us to entirely rule out this
possibility, such a contribution seems unlikely to be significant for
a number of reasons.

First of all, as mentioned, the hot halo properties of NGC~5746 are at
variance with those of other spirals where the halo emission can be
associated with SN outflows. In particular, Fig.~\ref{fig,strickland}
suggests that either mechanisms other than stellar processes are at
work in generating the NGC~5746 halo, or, rather unlikely, the SN
thermalization efficiency in the quiescent NGC~5746 is an order of
magnitude higher than for the comparison sample shown in the figure,
which includes actively star-forming spirals such as the prototypical
starburst outflow galaxies M82 and NGC~253. The low metallicity
derived for the halo gas also argues against a disk origin, although
the robustness of this measurement can be questioned.

A second argument against stellar activity being responsible for the
observed emission around NGC~5746 bases itself on the possibility of
gas being blown out of the disk by SN activity. The large depth of the
gravitational potential of NGC~5746 would be one factor which could
inhibit stellar outflows from the disk.  In practice, however, the
distribution and pressure of the interstellar medium are usually more
important factors for the possibility of blow-out than the depth of
the gravitational potential (e.g.\ \citealt{macl1988,stri2000}). In a
study of extraplanar X-ray emission detected around spiral galaxies
covering a range of morphologies and star formation rates,
\citet{stri2004} derived a criterion for gas blow-out based on the
observed properties of the interstellar medium in typical spirals.
They find on theoretical grounds that in order to have a collection of
near-simultaneous SN expel gas out of the disk, the requirement
$F_{{\rm SN,FIR,}D_{25}} > 25$~SN~Myr$^{-1}$~kpc$^{-2}$ must be
satisfied.  In their notation, $F_{{\rm SN,FIR,}D_{25}}$ is the rate
${\cal R}_{\rm SN} = 0.2L_{\rm IR}/10^{11}$~L$_{\sun}$ of
core-collapse SN per disk ``area'' $D_{25}^2$, again assuming $L_{\rm
  IR}$ to be a proxy of the total star formation rate. Indeed, for the
seven spirals in their sample with detected extraplanar X-ray
emission, they find $F_{{\rm SN,FIR,}D_{25}} >
40$~SN~Myr$^{-1}$~kpc$^{-2}$.  By contrast, using an approach
identical to that of \citet{stri2004}, we obtain $L_{\rm IR} = 7\times
10^9$~L$_{\sun}$ and $F_{{\rm SN,FIR,}D_{25}} \approx 4$ SN Myr$^{-1}$
kpc$^{-2}$ for NGC~5746. Thus, the idea that star formation activity
in NGC~5746 can blow out large quantities of gas to distances of $\sim
20$ kpc above the plane is in clear conflict with the results of
\citet{stri2004}. This result, of course, neglects the contribution
from type~Ia SNe, the rate of which may depend more sensitively on the
total stellar mass than on the disk star formation rate. From the $K$-
and $B$-band magnitudes of NGC~5746, a type Ia SN rate of $\approx
0.01$~yr$^{-1}$ can be estimated, using the prescription of
\citet{mann2005}. This would imply $F_{{\rm SN,FIR,}D_{25}}\approx
7$~SN~Myr$^{-1}$~kpc$^{-2}$ for the total SN rate (equivalent to
0.02~SN~yr$^{-1}$ over the full disk), still well below the criterion
derived by \citet{stri2004}.

Although the present level of stellar activity in the disk is thus
unlikely to explain the X-ray emission around NGC~5746, the
possibility remains that the gas was deposited in the halo by previous
such activity. This would have had to happen sufficiently long ago to
allow any H$\alpha$ and radio halo to fade away and return standard
SFR indicators such as $L_{\rm FIR}$ of the disk to quiescent values.
Indeed, it was argued in Paper~I that the energy release required to
set up the halo would involve the formation of 5--10$\times
10^9$~M$_\odot$ of stars, and that the age of such a hypothetical
starburst would have to be at least 1--2~Gyr in order to comply with
the present optical colours of NGC~5746. On the other hand, the
derived hot gas properties translate into a {\em mean} cooling time at
$D_{25}$ of $t_{\rm cool} = 3.9^{+2.0}_{-0.9} \times 10^9$~yr. This
sets a natural upper limit on the age of such a starburst.

From the observed correlation between far-infrared disk luminosity and
the X-ray extent of a hot halo generated by an active starburst
\citep{grim2005}, one would expect a characteristic extent of the
extraplanar X-ray emission around NGC~5746 of only 1--2~kpc. This is
in itself an argument against {\em ongoing} star formation activity
being responsible for the NGC~5746 halo. If deposited in the halo by a
starburst, the observed extent of halo gas would suggest a peak
$L_{\rm FIR}$ of $\sim 100$ times its present value of $L_{\rm
  FIR}=3.5\times 10^9$~L$_\odot$, equivalent to an estimated SFR of
order 50--100~M$_\odot$~yr$^{-1}$ (using the relations described in
\S~\ref{sec,sample}). A starburst of this magnitude would have placed
NGC~5746 among ultraluminous infrared galaxies (ULIRGs) rather than
``normal'' starburst \citep{grim2005}.  The cause of such a powerful
starburst occurring within the last few Gyr is not clear, however;
ULIRGs seem to be the products of major mergers of gas-rich galaxies
(mass ratio less than 3:1; \citealt{dasy2006}), but NGC~5746 is
isolated and does not show any immediate evidence for previous violent
interactions with other galaxies.

In summary, we cannot entirely rule out a stellar origin for the
extraplanar X-ray emission, but comparison to other spirals with and
without such emission indicates that the halo of NGC~5746 does not fit
well into such a scenario.

\subsection{AGN Outflows}

Outflows of gas could alternatively be produced by AGN activity.  The
brightest central X-ray point source inside $D_{25}$, at
$(\alpha,\delta)_{2000} = (14^h44^m56\fs 0, 01\degr
57\arcmin18\arcsec)$, exhibits a total of $\sim 350$ net counts
(0.3--5~keV).  Fitting a doubly absorbed power-law to its spectrum
produces a good fit ($\chi^2_\nu = 0.89$), with $\Gamma =
1.38^{+0.29}_{-0.22}$, and $N_{\rm H} = 7.7^{+3.5}_{-3.4}\times
10^{21}$~cm$^{-2}$ for the intrinsic absorption. While this is
consistent with a typical AGN spectrum, the absorption-corrected
0.3--5 keV luminosity of $1.9_{-0.7}^{+1.2} \times
10^{40}$~erg~s$^{-1}$ suggests very moderate activity.  Given the
distance to NGC~5746, it is furthermore possible that the source in
fact remains unresolved in our X-ray data. The {\em Chandra} spatial
resolution corresponds to a physical extent of $\ga 75$~pc at this
distance, so the derived luminosity should probably be viewed as an
upper limit. There is also no evidence for any significant nuclear
radio source in the VLA data (cf.\ Fig.~\ref{fig,overlays}).
Moreover, any associated AGN outflow perpendicular to the disk would
be highly collimated, and there is no indication that this is the case
(Fig.~\ref{fig,overlays}). A 2--7~keV image, in which any AGN should
stand out more prominently than in Fig.~\ref{fig,overlays}, confirms
the lack of spatial substructure in the hard X-ray emission and shows
that this emission is confined to the very central regions of the
disk. The stellar kinematics in the nuclear region also shows that any
central black hole must be of rather modest mass \citep{bowe1993}.

Finally, gas in an AGN outflow would also be unlikely to have attained
thermal equilibrium and so would probably not be well described by a
thermal plasma model, in contrast to observations. Indeed, AGN
outflows in galaxy clusters (where they can be more easily detected
due to the influence on the surrounding intracluster medium) seem to
be associated with bubbles of relativistic non-thermal radio plasma
(e.g.\ \citealt{birz2004}), rather than with soft X-ray emitting
material.  In fact, these radio bubbles typically coincide with
``cavities'' in the surrounding intracluster X-ray gas, and in the
well-studied case of the Perseus cluster are consistent with
containing {\em no} detectable thermal X-ray gas, even in very deep
X-ray observations (J.\ Sanders, private communication). The detection
of an AGN-generated X-ray halo around NGC~5746 in these {\em Chandra}
data would therefore be a major surprise.  Hence, as also concluded in
Paper~I, it seems safe to exclude the possibility that recent or
previous AGN activity should be responsible for the extraplanar X-ray
emission in this case.

\subsection{Accretion of Intergalactic Gas}\label{sec,accretion}

The only obvious remaining explanation for the X-ray emission around
NGC~5746 is the radiative cooling of hot gas residing in an extended
gaseous halo which has not been generated by processes in the disk.
Since such externally accreted gas is expected to have low
metallicities, whereas disk outflows could have (super)solar
abundances, a decisive way to test the accretion scenario against an
outflow origin for the hot gas would be a robust estimate of the gas
metallicity. This could be achieved with deep X-ray spectroscopy, but
is unfortunately not feasible with the present data.  There are,
however, a number of arguments which strengthen the interpretation
that the X-ray emission represents externally accreted gas.

(i) The low metallicity derived for the halo gas, $Z \approx
0.1$~Z$_\odot$, indicates an external origin, although, as mentioned,
this result could be affected by the Fe bias. Taken at face value, the
result is in good agreement with that found for gas presumably
accreting onto the Milky Way \citep{wakk1999}, and also corresponds to
the value derived for the low-redshift intergalactic medium
\citep{danf2006}.

(ii) The temperature of halo gas is consistent with the expected
virial temperature of the underlying dark matter halo, suggesting that
the gas has been gravitationally heated. This is generally not the
case for X-ray gas within (stellar) outflows, which show temperatures
that are largely independent of the stellar (e.g.\ \citealt{rasm2004b}
or total (see \S~\ref{sec,comparison}) galactic mass. Of course, the
agreement between $T_{\rm vir}$ and the observed temperature for
NGC~5746 could be a coincidence, but:

(iii) Not only the temperature, but also the X-ray luminosity, surface
brightness profile, and total hot gas mass of the NGC~5746 halo are
matched by predictions of cosmological simulations involving infall of
hot gas onto spirals (see \S~\ref{sec,comparison}). Again, a good
match between simulations of infall and disk outflows would generally
not be expected, and, indeed, is not observed. If the hot gas
surrounding NGC~5746 originated within the disk, the agreement with
our simulations in \S~\ref{sec,comparison} would be a spectacular and,
to our knowledge, unique coincidence.

Although a direct confirmation of the accretion scenario would require
more sensitive X-ray data, it appears to be the most plausible
explanation in light of the above discussion.  When interpreted in
this way, the extraplanar X-ray emission of NGC~5746 thus represents
the first tentative detection of the hot halos of externally accreted
gas in which massive spiral galaxies are believed to be embedded. One
should note, then, that simulations show such halo gas to display a
range of temperatures (albeit a fairly narrow one outside the disk;
see \S~\ref{sec,comparison}), so the model fits of
Table~\ref{tab,spec} should be considered a convenient means of
parametrizing the halo X-ray properties rather than a complete
description of the physical state of hot halo gas.

\subsection{Is the Halo Gas Currently Inflowing?}\label{sec,infall}

In the accretion scenario discussed above, halo gas is expected to be
actively accreting onto the disk due to the decreased pressure support
of the rapidly cooling gas close to the disk. The question of whether
the observed gas is really falling in at present naturally bears some
connection to the issues discussed above.  As mentioned, the mean
cooling time at $D_{25}$ is only $\approx 4$~Gyr, and any thermal
instabilities in the gas would act to locally reduce this value,
likely leading to the condensation of clouds of colder material which
can accrete onto the disk \citep{kauf2006,somm2006}.  Therefore, if
cooling and the ensuing inflow of halo gas is not counteracted by
feedback, the X-ray emitting material should currently be accreting
onto the disk.  However, although the halo itself may be unlikely to
result from SN/AGN activity, it is conceivable that the energy
released by such processes is currently preventing halo gas from
actually inflowing (see e.g.\ \citealt{binn2004}), either through
mechanical $pdV$ work done on this gas or through re-heating of gas
which has cooled to $T\la 10^6$~K.  Unfortunately, the present data do
not allow us to directly assess the validity of this scenario, but
some insight can be gained from more indirect arguments. If assuming
that the spatial distribution of energy input to the halo roughly
matches that of radiative losses from the halo, an upper limit to the
energy required to offset cooling and gas infall would simply be the
halo luminosity, $L_{\rm X }\approx 7\times 10^{39}$~erg~s$^{-1}$.

In principle, the mechanical output generated by an AGN could provide
this power. Strong radio sources associated with nuclear activity in
elliptical galaxies are emerging as prime candidates for retarding and
even quenching cooling flows on a variety of scales. This probably
occurs via a feedback-driven activity cycle of duration $\sim
10^8$~yr, and could be effective both in individual ellipticals and in
the cores of groups and clusters (e.g.\
\citealt{fabi2003,voit2005,best2006}).  However, there is little
evidence from Fig.~\ref{fig,overlays} for current or even past strong
radio activity in NGC~5746 (i.e.\ a bright nuclear radio source, radio
outflows from the disk, and/or extended radio emission some distance
from the disk as a result of previous activity). This suggests that
any AGN may currently be in a quiescent phase of its activity cycle,
but it also precludes any direct estimate of the energy output of an
AGN in its active phase.

Nevertheless, an estimate of the maximum AGN energy output in NGC~5746
may be obtained from the results of \cite{birz2004}. These authors
studied a sample of 18 galaxy systems ranging from rich clusters down
to a single elliptical galaxy, in which cavities in the surrounding
X-ray gas can be associated with outflows generated by current or past
activity of a radio source in the central galaxy. They found a
correlation between the radio luminosity of this central source and
estimates of the mechanical luminosity $L_{\rm mech} = W/t$ of the
cavities. Here $W$ is the $p\mbox{d}V$ work done by the cavity on the
surrounding X-ray gas as it rises buoyantly through this gas, and $t$
is the age of the cavity.  Though the derived relation is subject to
substantial scatter, it remains useful for providing a rough estimate
of the mechanical energy released by the central radio source in
NGC~5746 in any active phase. The 1.4~GHz continuum flux of this
source of 14.9~mJy \citep{cond1998} would correspond to a total radio
luminosity of $\sim 1\times 10^{38}$~erg~s$^{-1}$, calculated using
the prescription of \citet{birz2004} and assuming a spectral index
$\alpha=1$.  In turn, the relation derived by \citet{birz2004} would
then suggest $L_{\rm mech}$ of order $10^{41}$~erg~s$^{-1}$ which
could be sufficient to counteract infall.

However, as also noted by \citet{birz2004}, this estimate could suffer
from considerable systematic uncertainties, and it is moreover unclear
whether their results are representative of the overall efficiency
with which AGN mechanical energy is transferred to its surroundings.
For example, the result is based on a sample for which X-ray cavities
{\em can} be detected and which does not include any spiral galaxies.
The only safe conclusion, it seems, is that an AGN in an active phase
(assuming AGN in isolated spiral galaxies have activity cycles as seen
for central cluster galaxies, although there is little support for
this assumption at present) could potentially halt cooling and infall
of halo gas, but that there is no significant evidence from X-ray or
radio data of any ongoing or recent activity of this type.  Hence, our
observations do not allow a direct test of the hypothesis put forward
by \cite{binn2004}, that once a hot gaseous halo has been built up
around a galaxy, its gas does not cool but rather remains thermostated
at $T\approx T_{\rm vir}$ due to energy injection from an AGN (and
stars). Although significant AGN activity does not seem to be taking
place at present in NGC~5746, the existing radio data probably lack
the sensitivity to allow an unambiguous test for such activity in the
recent past. This is certainly true for the X-ray data, leaving
Binney's hypothesis a viable possibility.

Supernova outflows provide an alternative means of preventing infall
of halo gas. While the total SN rate of NGC~5746 of $\sim
0.02$~yr$^{-1}$ (\S~\ref{sec,stars}) is probably insufficient to blow
gas out of the disk, as discussed above, it still corresponds to a
mechanical energy release of $\approx 1\times 10^{42}$~erg~s$^{-1}$
for a typical SN output of $10^{51}$~erg. Simulations suggest that,
under normal interstellar conditions, most ($\sim 90$~per~cent) of
this energy is radiated away \citep{thor1998}; in fact, this would be
consistent with the diffuse disk luminosity of NGC~5746 if assuming an
intrinsic absorption in the disk equal to that derived for the central
X-ray point source. Of the remaining energy, a small fraction is
expected to thermalize in the interstellar medium of the disk, leaving
a mechanical luminosity which could still be an order of magnitude
larger than the halo $L_{\rm X}$.  Although insufficient to blow gas
out of the disk and hence entirely inhibit halo gas infall, it will,
in particular in combination with any nuclear activity, act to reduce
the infall rate and perhaps also reheat some of the already cooled
accreting gas, making it at least temporarily unavailable for star
formation. In practice, this implies that the derived mass accretion
rate $\dot M$ of external halo gas must be viewed as an upper limit
under the given assumptions.

Future tests, involving deeper X-ray and far-ultraviolet observations,
could shed light on these issues.  For example, if much of the halo
gas is held in place by activity in the disk, one would expect
enriched material from the disk to have mixed with halo gas, so a more
robust X-ray metallicity constraint would be highly useful. To test if
cooling does occur outside the disk, one could look for evidence of a
temperature gradient in the halo X-ray gas (although, as discussed in
\S~\ref{sec,comparison}, the absence of a $T$-gradient outside the
disk does not necessarily imply that gas is not cooling rapidly even
further in). Finally, one could test whether gas is cooling below
X-ray temperatures using {\em FUSE} observations of O{\sc vi} at a
range of distances from the disk, although the expected low metal
content of halo gas may render such observations difficult.

Summarizing, there is no direct observational evidence for any AGN or
SN activity at levels sufficient to completely offset cooling and
infall of halo gas in NGC~5746, and indirect arguments indicate that
the required power for this to occur is not currently available.
Consequently, given the derived cooling time of halo gas, some of this
gas should currently be accreting onto the disk, acting as a fresh
supply of material for continued star formation. If so, the disk of
this massive spiral is still being built up at present.

\section{Comparison to simulations of galaxy formation}\label{sec,comparison}

In the accretion scenario, halo properties can be predicted in detail
by cosmological simulations. Hence, it is natural to compare the
results obtained for both our galaxies to numerical work.  To this
end, we have employed cosmological TreeSPH simulations of galaxy
formation and evolution in a flat $\Lambda$CDM cosmology.  The code,
described in detail in \citet{somm2005b} and \citet{rome2006},
incorporates energetic stellar feedback taking the form of
starburst-driven galactic winds.  With respect to our earlier
simulations used by \citet{toft2002} to predict halo properties (as
discussed in \S~\ref{sec,intro} and \S~\ref{sec,sample}), the present
simulations have been updated to incorporate chemical evolution of the
gas, including non-instantaneous recycling, and are evolved in time
according to the entropy equation solving scheme of \citet{spri2002},
rather than being based on the thermal energy equation. The latter
improvement provides increased numerical accuracy in lower-resolution
regions. In addition, the present simulations have all been run with
at least eight times higher mass resolution. The highest resolution
runs of individual galaxies, at 512 times the resolution of the
simulations described by \citet{toft2002}, feature a total of $\sim
2\times 10^5$ gas and $\sim 1.5\times 10^5$ dark matter particles
inside (100 kpc)$^3$, a gas mass resolution of $1.2\times
10^4$~M$_{\sun}$, and a corresponding gravitational softening length
of $\sim 130$~pc.

Although the simulations include stellar feedback, optionally enhanced
to mimic AGN outflows, starburst-driven winds are only invoked at
early times ($z\ga 4$--5), in order to solve the angular momentum
problem, the missing satellites problem, and possibly other problems
related to the cold dark matter scenario (see, e.g.,
\citealt{somm2003}).  At $z\la 4$, the SN~II feedback is quite
moderate in our simulations; for a typical $v_c =220$~km~s$^{-1}$ disk
galaxy, the star formation rate at $z=0$ is $\sim
0.5-1$~M$_\odot$~yr$^{-1}$, similar to the modest star formation rate
of the Milky Way disk.  The feedback of the individual star particles
in the simulations is furthermore spatially and temporally
uncorrelated, implying that starburst winds do not easily develop at
low redshift (see also \citealt*{dahl2006}). Hence, a comparison
between the simulations and our two relatively quiescent galaxies
seems justified.

It is worth emphasizing that the feedback strength in the simulations
is calibrated to reproduce the {\em optical} properties of observed
galaxies, but that the X-ray emission is predicted without any
additional adjustable parameters. We find that neglecting feedback in
the simulations tends to decrease the present-day halo X-ray
luminosity, because less feedback implies that hot halo gas cools out
faster over cosmic time, leaving less hot gas in the halo at $z=0$
(the situation is qualitatively similar to the difference between
assuming fixed, primordial metallicity and $Z=0.3$~Z$_\odot$ for the
halo gas in the simulations, as discussed by \citealt{rasm2004a}). We
note in passing that this also implies that the predictions of
semi-analytical models of galaxy formation (which tend to over-predict
halo $L_{\rm X}$ with respect to observations) and those of our
simulations cannot be reconciled simply by neglecting feedback in the
simulations.

For the calculation of X-ray properties of the simulated galaxies, we
follow the procedure outlined by \citet{toft2002}, except that the
increased numerical resolution of the present simulations allows us to
bypass the smoothing over the individual gas particles in the
simulations. All simulation results have been derived inside a region
corresponding to the adopted NGC~5746 halo aperture (region
B$\setminus$E of Table~\ref{tab,spec}). A few comparisons between
NGC~5746 and results from these simulations were performed in Paper~I,
showing the simulation predictions to be in good agreement with the
observed X-ray surface brightness profile and integrated halo $L_{\rm
  X}$.  Below we will elaborate on these and other simulation results,
present additional comparisons to observed galaxies, and discuss the
implications for models for the formation of spiral galaxies.

\subsection{X-ray Luminosity, Temperature, and Surface Brightness}

Figure~\ref{fig,LxVc}$a$ displays results based on our recent
simulations described above, run at 8, 64, and 512 times the
resolution of the \citet{toft2002} simulations. Of the new
simulations, the one producing the galaxy with $v_c\approx
226$~km~s$^{-1}$ assumes a cosmic baryon fraction $f_b=0.10$, whereas
those of higher--$v_c$ galaxies assume $f_b=0.15$.  Also shown are
results for a subsample of the simulated galaxies of \citet{toft2002}.
These were all formed in simulations assuming either a $\Lambda$CDM or
$\Lambda$WDM cosmology of $\Omega_m=0.3$ and $\Omega_{\Lambda}=0.7$,
$f_b = 0.10$, and a (fixed) primordial chemical gas composition.  The
X-ray luminosity of these simulated galaxies have been corrected for
the error described in \citet{rasm2004a}.  The correlation of $L_{\rm
  X}$ with $v_c$ is seen to display notable scatter, most of which is
related to intrinsic differences in the accretion history of the
galaxies \citep{toft2002}.  In general, there seems to be good
agreement between the two simulation samples.  While numerical
resolution effects seem to play a limited role in establishing the
predicted $L_{\rm X}$--$v_c$ trend or its scatter, there is some
indication that increased resolution tends to lower the predicted halo
luminosity by a small amount.

\begin{figure*}
\begin{center}
\epsscale{1.0}
\plottwo{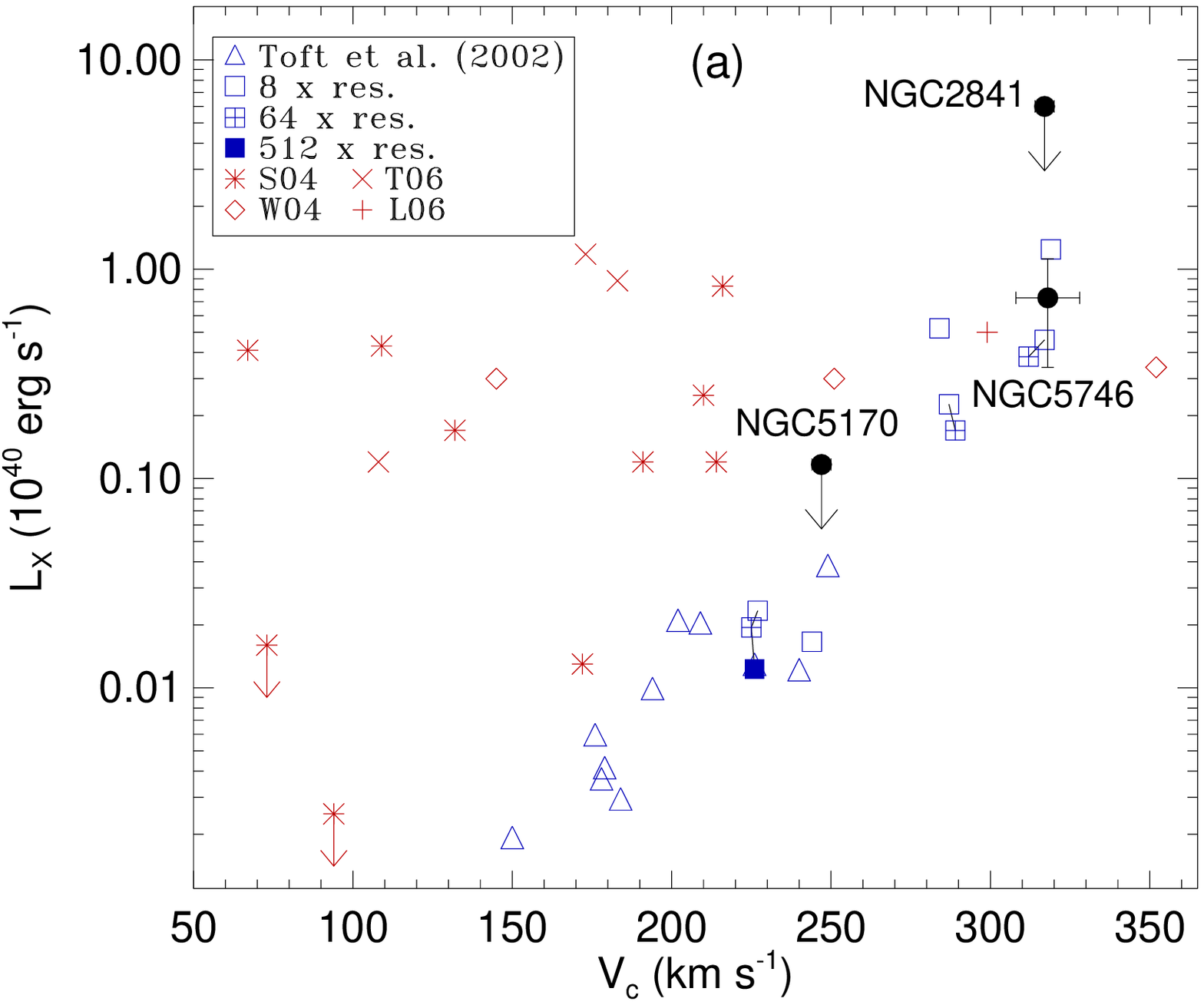}{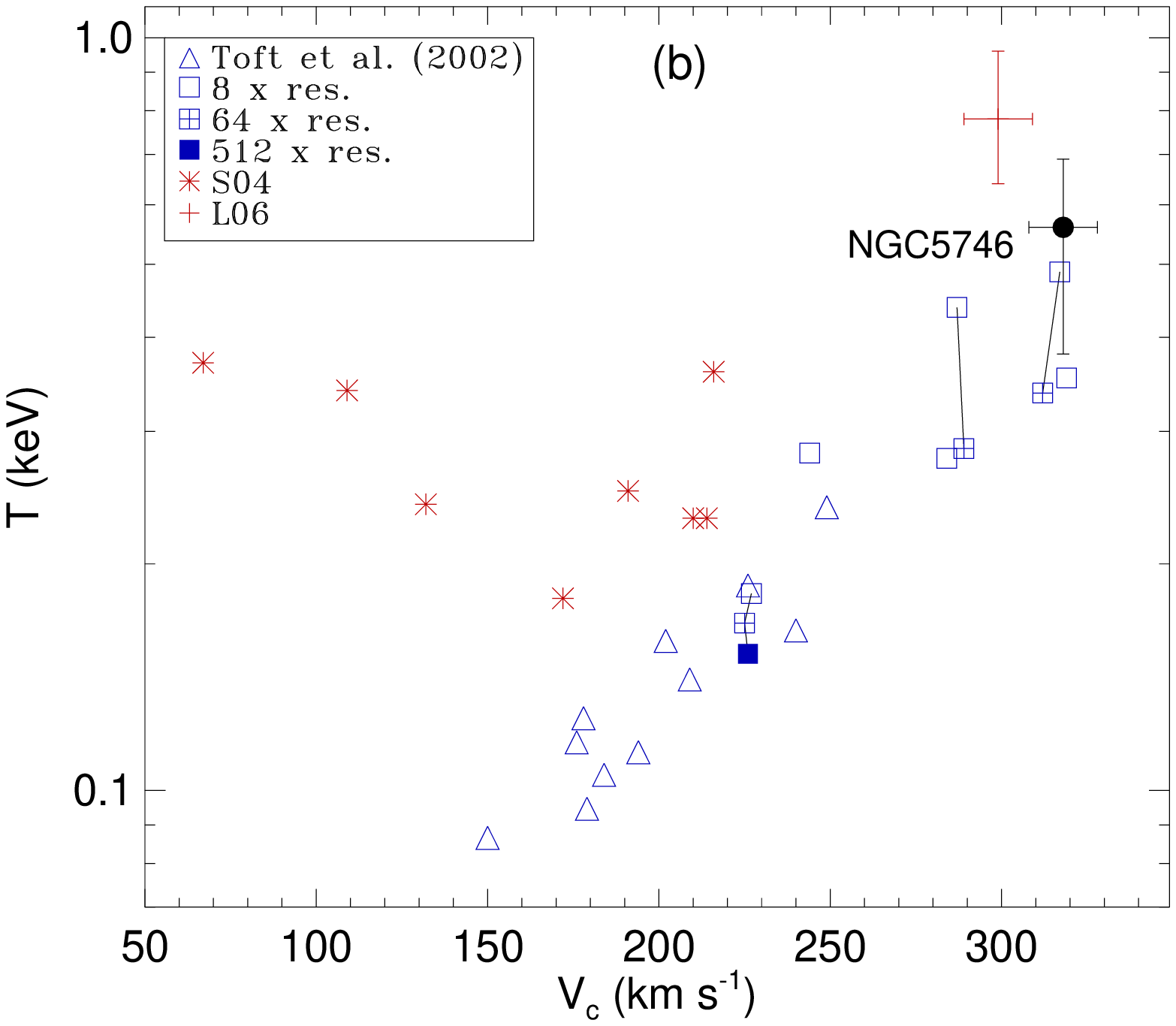}
\end{center}
\figcaption{(a) X-ray luminosity and (b) emission-weighted hot halo
  temperature versus $v_c$ for the simulated galaxies of Toft
  et~al.\ (2002; triangles) and for those in our new simulations
  (squares), all calculated assuming the NGC~5746 halo aperture.
  Simulations of the same galaxy run at different resolutions are
  connected by a line. Also shown are the available observational
  results, with symbols as in Fig.~\ref{fig,strickland}.
\label{fig,LxVc}}
\end{figure*}

Figure~\ref{fig,LxVc}$a$ also displays the derived constraints on the
halo luminosities of NGC~5170 and NGC~5746, and has been updated from
Paper~I to show the additional constraints obtained for the
extraplanar emission seen around other spirals, i.e.\ the comparison
sample also plotted in Fig.~\ref{fig,strickland}.  In these cases, the
halo luminosities have not necessarily been derived within the same
aperture as adopted for NGC~5746 and our simulated galaxies. For
example, the results of \citet{stri2004} were all derived at a
distance $|z|>2$~kpc from the disk midplane. If instead adopting this
aperture, the increase in $L_{\rm X}$ predicted for our simulated
galaxies is at the 10--40~per~cent level, with no systematic
dependence on $v_c$. We expect similar corrections for the apertures
used for the other observed samples. These changes are relatively
small, being within the scatter seen for the simulated galaxies and
the estimated uncertainties in $L_{\rm X}$ for NGC~5746. We have
therefore chosen not to correct for this effect in
Fig.~\ref{fig,LxVc}$a$, as the overall conclusions will clearly remain
unaffected by such a change.  We also show the strongest constraint
resulting from the study of \citet{bens2000}, the $3\sigma$ upper
limit to the halo $L_{\rm X}$ of NGC~2841, here converted to the
0.3--2~keV band assuming a {\em mekal} thermal plasma of $T\approx
0.4$~keV based on the ($T,v_c$) plot of \citet{toft2002}.  This result
was originally derived within a 5--18~arcmin annulus around the
galaxy, so the value has been corrected to the NGC~5746 aperture on
the basis of the mean difference in $L_{\rm X}$ between the two
apertures as predicted for the galaxies in our new simulations
(amounting to a factor $\sim 3$ increase in $L_{\rm X}$).  As can be
seen from the figure, the simulation results generally appear to be in
excellent agreement with those of our observed galaxies. In
particular, the theoretical models provide a good match to the derived
halo luminosity of NGC~5746, and their predictions can easily be
accommodated by the observational results for the (presumably mainly
feedback-generated) hot halos seen in other studies.

It is worth noting that the emission-weighted mean metallicity of the
X-ray gas in these simulations is very low, of order
$10^{-3}$~Z$_{\sun}$, predicting the inflowing gas to be essentially
primordial in composition. Those simulations of \cite{toft2002} that
assume a fixed metallicity of 0.3~Z$_{\sun}$ would, for a given $v_c$,
predict present-day luminosities 3--4 times {\em lower} than those
shown in Fig.~\ref{fig,LxVc}$a$ (cf.\ the above discussion).  This would
still leave NGC~5746 within the dispersion in the predicted $L_{\rm
  X}$--$v_c$ trend, but the agreement would be considerably less
convincing. This may indicate that any hot gas accreted by spirals
today is nearly pristine; a similar result emerges when comparing
simulations of galaxies at high redshift to observational constraints
on the redshift evolution of the {\em integrated} X-ray luminosity of
normal spirals \citep{rasm2004a}.

Figure~\ref{fig,LxVc}$b$ shows the temperature of hot halo gas versus
$v_c$. Despite the large variation in accretion history among the
simulated galaxies, the simulations predict a remarkably tight
correlation between the two quantities. The prediction is consistent
with the value derived for NGC~5746 and with the expected virial
temperature of its dark matter halo, $T_{\rm vir}\approx 0.3$ keV. The
result for NGC~5746 is larger than $T_{\rm vir}$ but, at a level of
$1.4\sigma$, not significantly so.  A plausible explanation could be
related to the fact that the hot halo spectrum of NGC~5746 is
dominated by emission close to the disk, and, as demonstrated below,
the simulations predict that the temperature of hot halo gas rises
slightly above the virial temperature close to the disk. We note that
a halo gas temperature slightly exceeding the virial temperature of
the dark matter is also a generic feature of the hot halo models
considered by \citet{fuku2006}.

For comparison to our simulations, the corresponding data for the
\citet{stri2004} sample and the \citet{li2006} result have once again
been overplotted in Figure~\ref{fig,LxVc}$b$. Emission-weighted halo
temperatures have not been published for the samples of
\citet{wang2004} or \citet{tull2006a}, so these galaxies are not
included in this diagram. The lack of a systematic trend (in both
$L_{\rm X}$ and $T$) with $v_c$, and hence galactic mass, within these
samples is another argument against a gravitational origin for their
extraplanar emission. This interpretation gains support from the fact
that the hot halo temperatures of these galaxies consistently exceed
the virial temperatures expected from the circular velocity of the
disk and are furthermore invariant with respect to $L_{\rm X}$ over
four orders of magnitude in SFR \citep{grim2005}. Gas is almost
certainly outflowing rather than infalling in these cases.

It is also worth emphasizing that our non-detection of any hot halo
surrounding NGC~5170 is consistent with the simulation results,
assuming its luminosity and temperature within the relevant aperture
can be reliably predicted from Fig.~\ref{fig,LxVc}. For an estimated
halo luminosity and temperature of $L_X\approx 10^{39}$~erg~s$^{-1}$
and $T\approx 0.2$~keV, respectively, we predict only $\sim 15$ net
halo counts in 0.3--2~keV, well within the Poisson noise of the
background. This supports our use of the NGC~5170 data in
\S~\ref{sec,results} as a testbed for our analysis procedure.

In Fig.~\ref{fig,surfbright3}$a$, we plot surface brightness profiles
of the simulated galaxies, extracted within the same physical aperture
as for NGC~5746, and co-added on either side of the disk as in
Fig.~\ref{fig,surfbright2}. In order to include the result for
NGC~5746 in this diagram, the background subtracted count rates for
NGC~5746 were converted into unabsorbed fluxes assuming a {\em mekal}
plasma of the best-fitting halo parameters (Table~\ref{tab,spec}) and
the Galactic value of $N_{\rm H}$.  For this conversion, a 30~per~cent
systematic uncertainty was added in quadrature to the statistical
errors shown in Fig.~\ref{fig,surfbright2}, on the basis of the
derived errors on spectral parameters.  The simulations of galaxies
with $v_c \approx 300$~km~s$^{-1}$ seem to produce results consistent
with the observed NGC~5746 profile for the adopted aperture.  To give
an appreciation of the effects of numerical resolution, results for
the $v_c\approx 226$~km~s$^{-1}$ simulated galaxy also shown in
Fig.~\ref{fig,LxVc} are overplotted.  There is some indication that
the predicted profile flattens for higher resolutions, but not to an
extent that would compromise the above conclusion.  Furthermore,
simulation results for the innermost bin cannot be compared directly
to observations, because much of the halo emission in this region
would be absorbed by neutral gas in the disk of an edge-on galaxy and
would moreover suffer contamination from unrelated emission associated
with hot gas in the disk.

\begin{figure*}
\begin{center}
\epsscale{1.1}
\plottwo{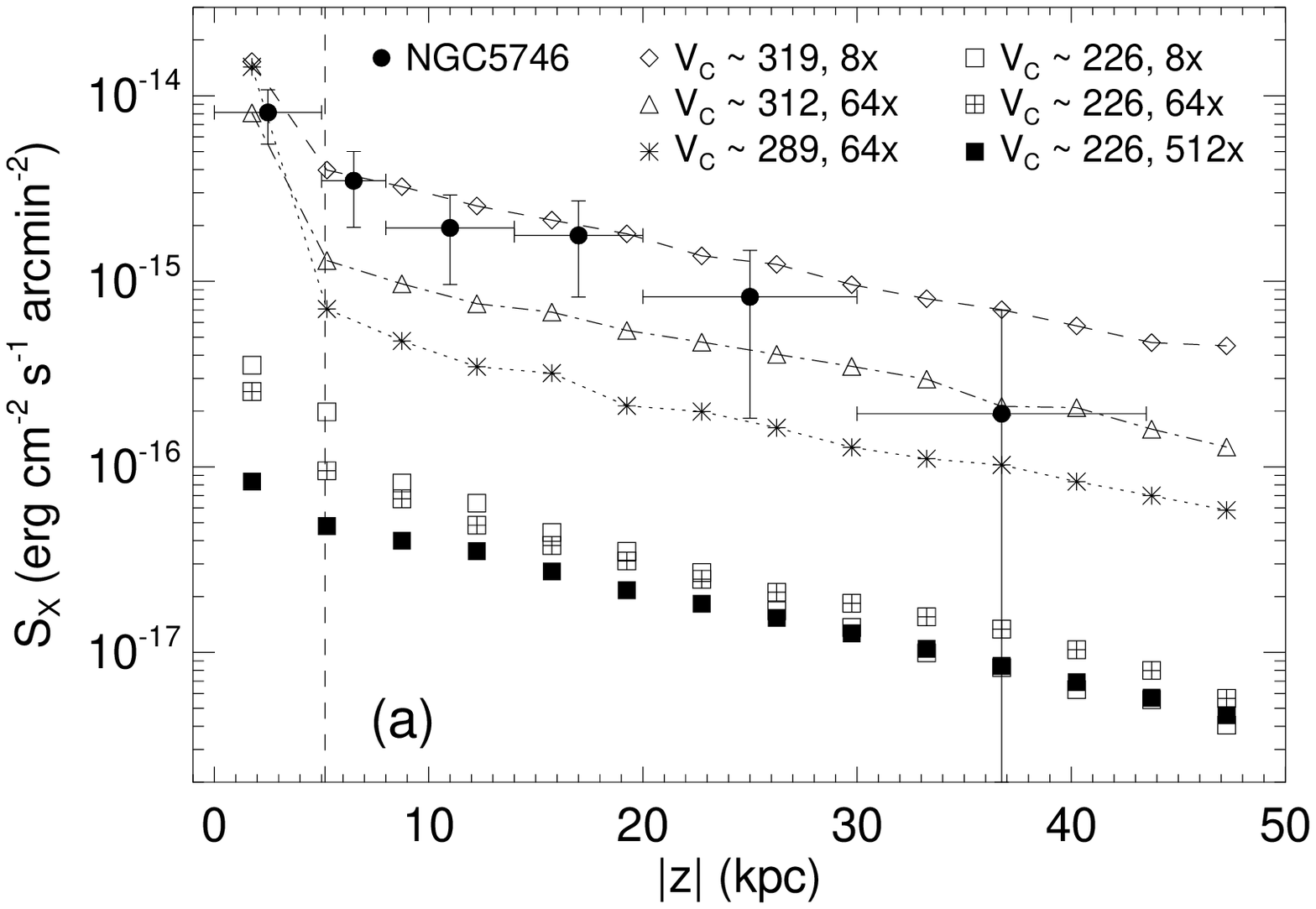}{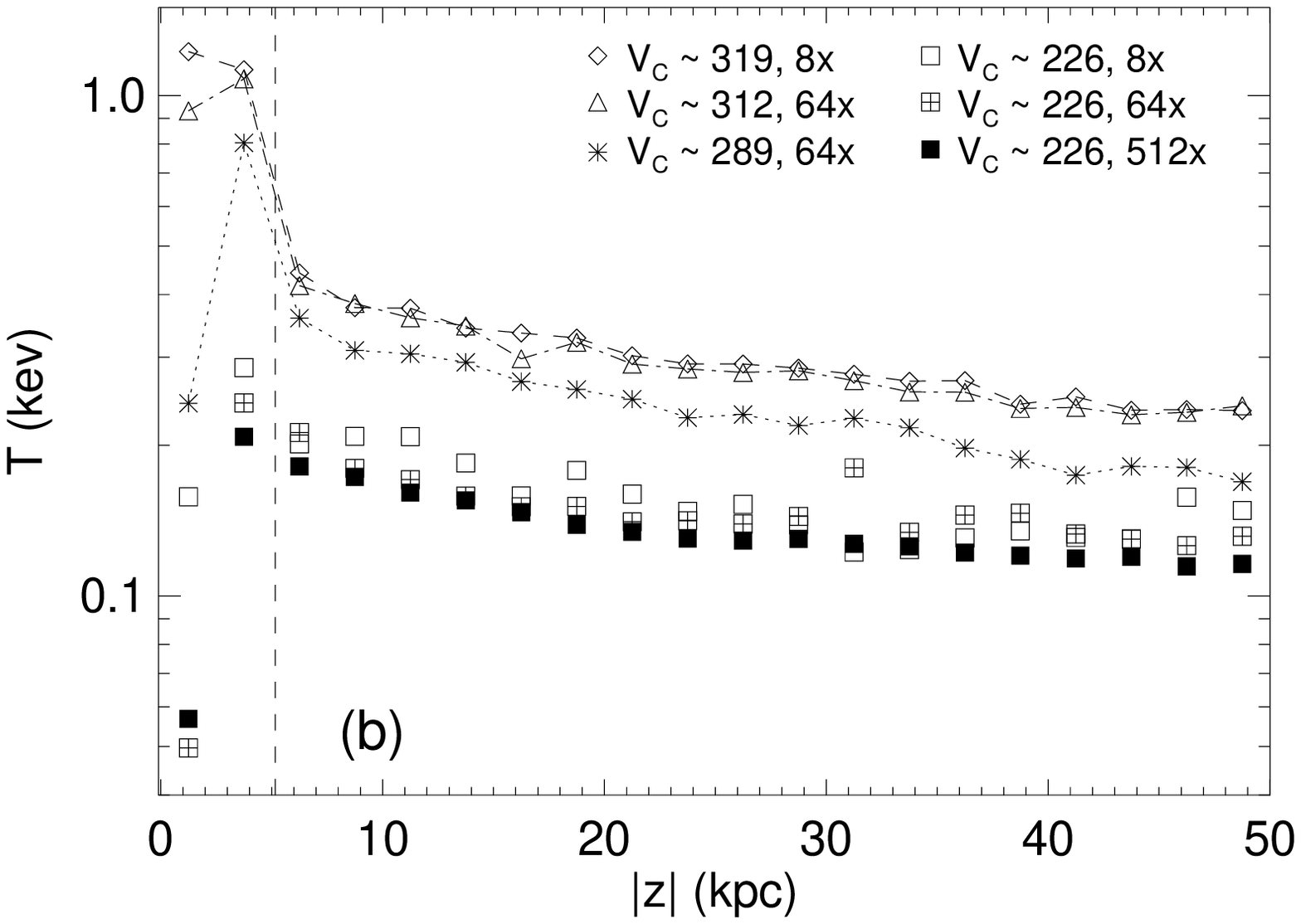}
\end{center}
\figcaption{(a) Derived surface brightness profile for NGC~5746 along
  with those of simulated galaxies in the relevant mass range, all
  computed inside a region corresponding to region~A of
  Fig.~\ref{fig,overlays}. Dashed vertical line marks the extent of
  the $D_{25}$ minor axis of NGC~5746. (b) The corresponding
  emission-weighted temperature profiles of the simulated galaxies,
  with symbols and lines as in the left figure.
\label{fig,surfbright3}}
\end{figure*}

The hot gas temperature of the simulated galaxies, plotted in
Fig.~\ref{fig,surfbright3}$b$, is seen to occupy a range of values,
which is, however, reasonably narrow outside the 'disk'. Well outside
the disk, the predicted halo temperature is in excellent agreement
with the virial temperature expected for the simulated galaxies (i.e.\
$T\approx 0.3$~keV for a galaxy with $v_c\approx 300$~km~s$^{-1}$).
Closer to the disk, $T$ increases gently and then rises steeply in the
innermost $\sim 5$~kpc above the disk midplane, followed by a strong
decline as the gas eventually reaches a density which enforces rapid
cool-out onto the disk. The reasonably uniform temperature
distribution predicted outside the disk suggests that our usage of a
single-temperature model for the description of the halo spectrum is
supported on a physical basis, and that the Fe bias may not be
significantly affecting our metallicity measurement.  However, it also
indicates that, although gas is cooling rapidly in the central halo
regions, the gas outside the disk could be near--isothermal.  One
implication is that X-ray observations of extraplanar gas around
spirals may not be able to directly test whether this gas is indeed
cooling out of the X-ray phase, as much of the cooling occurs very
close to the disk where contamination from other X-ray sources,
diffuse and point-like, could be substantial.

\subsection{Hot Gas Mass and Accretion Rates}

The amount of hot X-ray gas surrounding the simulated galaxies depends
not only on the mass of each galaxy's dark matter halo but also on the
detailed accretion history of this halo. Accretion histories can be
highly disparate even for isolated galaxies of comparable mass (e.g.\
\citealt{vand2002,rasm2004a}).  Any relation between the mass $M_{\rm
  gas}$ of hot halo gas and $v_c$ (or $L_{\rm X}$ and $v_c$) is thus
expected to display considerable intrinsic scatter, as indeed
evidenced by Fig.~\ref{fig,Mgas}$a$. In this plot we show the gas mass
of the simulated galaxies inside the volume assumed in the
corresponding calculation for NGC~5746 (\S~\ref{sec,halogas}), i.e.\
excluding the volume covered by $D_{25}$ of NGC~5746 out to 45~kpc
from the disk center along the line of sight. Given that the
simulations also include cold, non--X-ray emitting gas around the
galaxies and that $M_{\rm gas}$ is not an X-ray emission-weighted
quantity, a low-temperature cut of $T=0.1$ keV has been imposed on the
simulated galaxies in order to provide a fair comparison to
observations. This temperature is probably the lowest one at which gas
can be reliably detected by {\em Chandra} or {\em ROSAT} (see, for
example, the discussion in \citealt{rasm2004b}).  For completeness, we
also show the mean effect of this $T$--cut for four intervals in
$v_c$.  Simulation results are again seen to be in good agreement with
the observational constraints.

\begin{figure*}
\begin{center}
\epsscale{1.1}
\plottwo{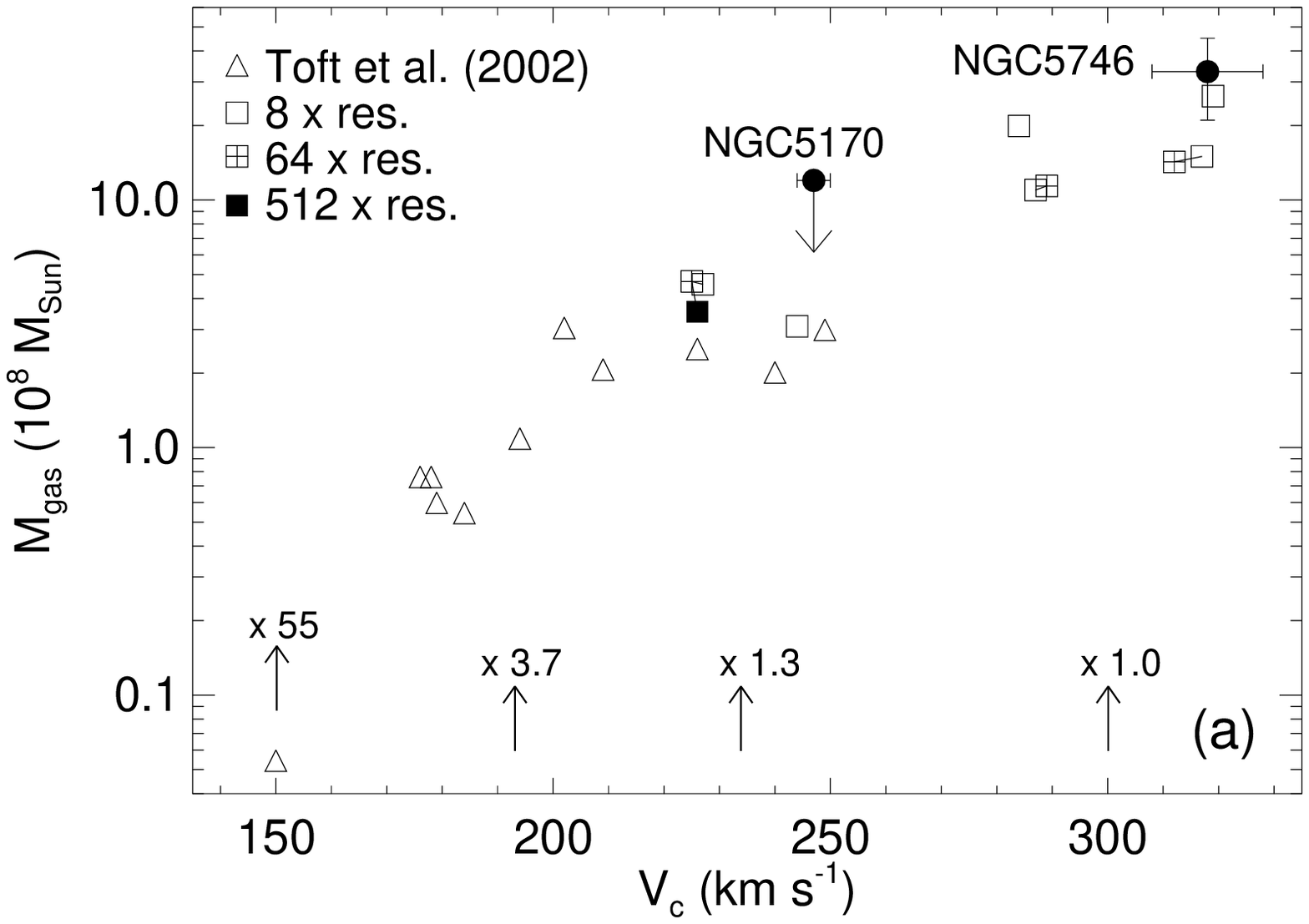}{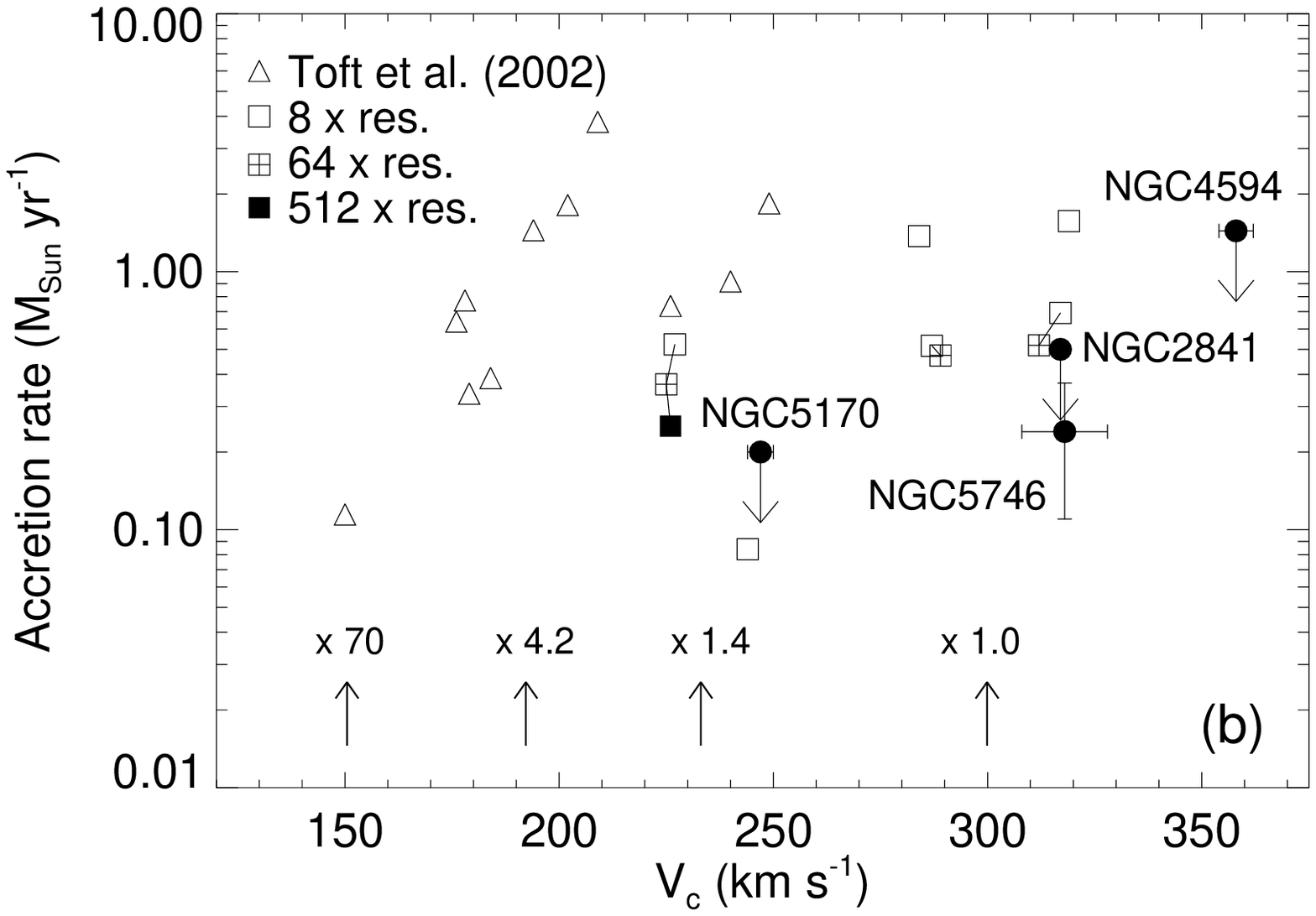}
\end{center}
\figcaption{(a) Mass of hot ($T\geq 0.1$ keV) gas as a function of
  $v_c$ within the NGC~5746 halo aperture.  (b) The corresponding hot
  gas accretion rate, $\dot M \sim M_{\rm gas}/t_{\rm cool}$.  Also
  shown are the upper limits estimated by \cite{bens2000} for NGC~2841
  and NGC~4594 (the 'Sombrero' galaxy).  Annotations at upward arrows
  give the mean factor of increase had all gas hotter than $10^5$~K in
  the simulations (within the aperture) been included in the estimate
  of $M_{\rm gas}$.  The numbers apply to the intervals $v_c < 160$ km
  s$^{-1}$, $170<v_c<210$ km s$^{-1}$, $210<v_c<255$ km s$^{-1}$, and
  $v_c > 255$ km s$^{-1}$.
\label{fig,Mgas}}
\end{figure*}

Fig.~\ref{fig,Mgas}$b$ shows the mass deposition rate $\dot M \approx
M_{\rm gas}/t_{\rm cool}$ of hot gas (which for all galaxies should be
viewed as an upper limit, cf.\ \S~\ref{sec,infall}).  Again, a
temperature cut at $T=0.1$~keV has been imposed on the simulations,
but $\dot M$ has otherwise been estimated in the same manner as for
the observed galaxies. The figure reveals substantial scatter at any
given $v_c$ and no clear systematic variation with this quantity. The
cool-out rate of NGC~5746 lies at the low end of the range predicted
by simulations, but there is also an indication that numerical
resolution effects might play a role, in the sense that increased
resolution tends to lower $\dot M$ within the adopted aperture.

The actual mass of hot halo gas which is cooling out around NGC~5746
would include any gas cooling within $D_{25}$. In the present data,
emission from such gas cannot be separated from diffuse disk emission
resulting from stellar activity.  Based on the simulated galaxies
covering the relevant mass range ($v_c > 255$~km~s$^{-1}$), we
estimate that accounting for cooling inside the disk region would
raise $\dot M$ by a factor $\xi \sim$~5--10, if assuming similar
levels of feedback in the simulated galaxies as for NGC~5746.  For the
derived value of $\langle \dot M\rangle$ (Table~ \ref{tab,halos}),
this would imply a total cool-out rate of $\sim
1$--2~M$_\odot$~yr$^{-1}$, in good agreement with the {\em
  IRAS}--based estimate of the disk star formation rate (SFR) of
NGC~5746, $0.8\pm 0.2$~M$_{\sun}$~yr$^{-1}$ (\S~\ref{sec,sample}).
While only an indicative result, this suggests that the galaxy is
currently forming stars at a rate corresponding to that at which gas
can be supplied for this process from the hot halo.  For NGC~5170, the
derived limit to the hot gas cooling rate, $\dot M_{\rm cool} <
0.2$~M$_{\sun}$~yr$^{-1}$, is a somewhat more model--dependent result
than that of NGC~5746, as it is based on an assumed value of the halo
temperature. The derived limit is nonetheless consistent with the disk
SFR estimate of $0.3\pm 0.1$~M$_{\sun}$~yr$^{-1}$, even without
accounting for the gas cooling out within the optical disk.

Can the disk of NGC~5746 have been built up by the accretion of hot
halo gas over cosmic time? Since stars are dominating the mass of
visible matter in the disks of spiral galaxies, the total baryonic
disk mass of NGC~5746 can be estimated from its $K$- and $B$-band
luminosities using the prescription of \citet{mann2005}.  This yields
a baryonic mass of $3.0\times 10^{11}$~M$_\odot$.  In order to
estimate the total mass $M_{\rm acc}$ of hot gas accreted by NGC~5746
since a given redshift, we can extrapolate the present accretion rate
by assuming
\begin{equation}\label{eq,mdot}
  \mbox{log}\dot M(z) \approx 0.6z + C.
\end{equation}
This follows the redshift evolution found by \citet{rasm2004a} for the
accretion of hot ($T\ga 3\times 10^5$~K) gas by spirals formed in
cosmological simulations. The relation was derived for simulated
galaxies with present-day masses similar to that of the Milky Way, and
applies to the redshift interval $z=0-2$, prior to which the galactic
disks themselves are not well-defined in the simulations.  For
NGC~5746, we have an upper limit (for $\xi=10$) of $\dot M(z=0)
\lesssim 4$~M$_{\sun}$~yr$^{-1}$.  Choosing the constant $C$ such as
to match this present upper limit, one then arrives at a corresponding
upper limit to the total mass of hot gas accreted by NGC~5746 since
$z=2$ of $M_{\rm acc}\lesssim 1.2\times 10^{11}$~M$_{\sun}$, which is
somewhat below, yet comparable to, the present baryonic mass in the
disk of NGC~5746.

However, this extrapolation is based on simulation results for Milky
Way--sized galaxies and could be subject to considerable systematic
uncertainty. On the one hand, some accretion would have taken place
before $z=2$, during a period in which the star formation activity and
the growth of a central supermassive black hole would have been much
more vigorous and could have counteracted gas accretion to a
significant extent.  On the other hand, given its deeper gravitational
potential, it is conceivable that accretion was initiated earlier for
NGC~5746 than for typical Milky Way--like galaxies (on which
Eq.~\ref{eq,mdot} is based) and would have been more dramatic at high
redshifts than for such galaxies. A more pronounced redshift evolution
of the cooling rate than the one adopted here can, in fact, be
accommodated within the constraints on the redshift evolution of the
integrated X-ray luminosity of spirals derived from {\em Chandra} Deep
Field data \citep{rasm2004a}.

Consequently, the current observational evidence is consistent with
the notion that a sizable fraction of the disk mass of NGC~5746 has
been formed by the accretion of hot X-ray emitting halo gas. Although
this is in encouraging agreement with recent cosmological models for
the formation of massive spirals, we acknowledge that this conclusion
is clearly a model-dependent result. Moreover,
Equation~(\ref{eq,mdot}) does not account for the accretion of {\em
  cold}, non--X-ray emitting gas, the addition of which would likely
bring the extrapolation into even better agreement with the observed
disk mass.  It is therefore clearly premature to exclude the
possibility that a significant fraction of the infalling gas around
NGC~5746 was either never or only briefly heated to X-ray
temperatures. In that case, accretion may, at least partly, have
proceeded much as in the 'cold mode' accretion scenarios originally
suggested by \citet{binn1977} and later followed up by various authors
\citep{birn2003,binn2004,kere2005,somm2005a}.

\section{Summary}\label{sec,summary}

Our {\em Chandra} data of the two massive edge-on disk galaxies
NGC~5170 and NGC~5746 have revealed a $5\sigma$ detection of
extraplanar X-ray emission around the more massive NGC~5746, extending
to at least $\sim 20$~kpc above the disk.  No detection is found
around NGC~5170 using the same analysis methods, which rules out the
possibility that the NGC~5746 result is due to instrumental artefacts
or details of our analysis. Both galaxies are isolated and show no
signs of significant nuclear or star formation activity. For NGC~5746,
this is also confirmed from our H$\alpha$ imaging of this galaxy,
leading to the conclusion that the diffuse X-ray emission surrounding
NGC~5746 is best explained as the signature of externally accreted gas
cooling radiatively in an extended hot halo.  The spectrum of this hot
halo gas is well described by a thermal plasma model of temperature
$\sim 0.5$~keV, with an indication that the gas metallicity is very
low, $Z \la 0.1$~Z$_\odot$, although deeper data would be needed to
confirm this.  The total 0.3--2~keV luminosity of the hot halo is
$\sim 7\times 10^{39}$~erg~s$^{-1}$, while the corresponding $3\sigma$
upper limit for the NGC~5170 halo is $\sim 3\times
10^{39}$~erg~s$^{-1}$.

We have performed a detailed comparison of the derived constraints on
halo X-ray properties of both galaxies to cosmological simulations of
galaxy formation and evolution which predict the existence of such a
halo. Very good overall agreement is found for the case of NGC~5746,
supporting our interpretation that the detected extraplanar X-ray
emission around this galaxy reflects the cooling of accreted hot gas
rather than being related to ongoing processes in the disk (such as
supernova-driven outflows of gas). The weaker limits obtained for the
non-detected halo around NGC~5170 are also easily consistent with
simulation predictions.  We also note that the results obtained for
all other galaxies with detected extraplanar emission are consistent
with our simulations.

Although the presence of accreted hot gas halos around spiral galaxies
has been hypothesized for almost 50~years, the observed X-ray emission
surrounding NGC~5746 seems to be the first, albeit tentative,
detection of such a halo.  This lends support to one of the
outstanding issues in many models of disk galaxy formation, namely the
assumption that extended reservoirs of accreted hot gas exist around
isolated high-mass disk galaxies, potentially supplying material for
the formation and growth of the baryonic component of such galaxies
even to the present day.  Moreover, it indicates that the "missing"
galactic baryons are, in fact, not missing, but mostly are constituted
by hot halo gas, as discussed by \citet{somm2006}.  However, we stress
that a decisive test of the accretion hypothesis for the gas
surrounding NGC~5746 would require more sensitive X-ray data; in
particular, deep X-ray spectroscopy would enable a robust metallicity
estimate for the halo gas, helping to distinguish between infall and
outflow models.

The estimated cooling rate of hot halo gas around NGC~5746 is found to
be consistent with predictions from simulations and with the current
star formation rate in the disk. We have estimated the total mass of
hot gas accreted by this system over cosmic time, indicating that a
significant fraction of the current disk mass can have been formed
through the infall of hot gas. This is consistent with many
semi-analytical models of the formation of massive spirals. It remains
a viable possibility, however, that a substantial fraction of the
present material in the disk has been accreted as cold ($T\lesssim
10^5$~K), non-X-ray emitting gas (such as the high-velocity clouds
seen around the Milky Way). We find that the present data do not allow
for a direct quantification of the relative importance of 'cold' and
'hot' accretion over the lifetime of the galaxy. More sensitive X-ray
observations of NGC~5746 and similar galaxies, coupled with results of
dedicated galaxy formation simulations, would be needed to settle this
issue.




\acknowledgments 

This work has made use of the HyperLeda and NASA/IPAC (NED)
extragalactic databases, and the Two Micron All Sky Survey database.
JR and KP acknowledge support from the Instrument Center for Danish
Astrophysics (IDA) during the initiation of this work. JR also
acknowledges the support of the European Community through a Marie
Curie Intra-European Fellowship.  JSL acknowledges support from the
Villum Kann Rasmussen Foundation and from Dark Cosmology Centre. ST
and LFO acknowledge support from the Danish Natural Sciences Research
Council. AJB acknowledges support from a Royal Society University
Research Fellowship and from the Gordon \& Betty Moore Foundation. RGB
acknowledges the support of a PPARC Senior Fellowship. We gratefully
acknowledge abundant access to the computing facilities provided by
the Danish Centre for Scientific Computing (DCSC), with which all
computations reported in this paper were performed.  Dark
Cosmology Centre is funded by the Danish National Research Foundation.\\


\end{document}